\documentclass
[aps,prl,superscriptaddress,reprint,showpacs,amssymb,amsmath,floatfix,citeautoscript,longbibliography]{revtex4-1}

\usepackage{graphicx} 
\usepackage{dcolumn} 
\usepackage{bm} 

\usepackage{ulem}
\usepackage[]{color}

\begin{document}

\title{A RIXS investigation of the crystal-field splitting of Sm$^{3+}$ in SmB$_6$}

\author{Andrea~Amorese}
 \affiliation{Institute of Physics II, University of Cologne, Z{\"u}lpicher Stra{\ss}e 77, 50937 Cologne, Germany}
  \affiliation{Max Planck Institute for Chemical Physics of Solids, N{\"o}thnitzer Stra{\ss}e 40, 01187 Dresden, Germany}
\author{Oliver~Stockert}
  \affiliation{Max Planck Institute for Chemical Physics of Solids, N{\"o}thnitzer Stra{\ss}e 40, 01187 Dresden, Germany}
\author{Kurt~Kummer}
  \affiliation{European Synchrotron Radiation Facility, 71 Avenue des Martyrs, CS40220, F-38043 Grenoble Cedex 9, France}
\author{Nicholas~B.~Brookes}
  \affiliation{European Synchrotron Radiation Facility, 71 Avenue des Martyrs, CS40220, F-38043 Grenoble Cedex 9, France}
\author{Dae-Jeong~Kim}
  \affiliation{Department of Physics and Astronomy, University of California, Irvine, CA 92697, USA}	
\author{Zachary~Fisk}
  \affiliation{Department of Physics and Astronomy, University of California, Irvine, CA 92697, USA}	
\author{Maurits~W.~Haverkort}
  \affiliation{Institute for Theoretical Physics, Heidelberg University, Philosophenweg 19, 69120 Heidelberg, Germany}
\author{Peter~Thalmeier}
	\affiliation{Max Planck Institute for Chemical Physics of Solids, N{\"o}thnitzer Stra{\ss}e 40, 01187 Dresden, Germany}
\author{Liu~Hao~Tjeng}
	\affiliation{Max Planck Institute for Chemical Physics of Solids, N{\"o}thnitzer Stra{\ss}e 40, 01187 Dresden, Germany}
\author{Andrea~Severing}
  \affiliation{Institute of Physics II, University of Cologne, Z{\"u}lpicher Stra{\ss}e 77, 50937 Cologne, Germany}
  \affiliation{Max Planck Institute for Chemical Physics of Solids, N{\"o}thnitzer Stra{\ss}e 40, 01187 Dresden, Germany}	

\date{\today}

\begin{abstract}
The crystal-field (CF) splitting of the $^6H_{5/2}$ Hund's rule ground state of Sm$^{3+}$ in the strongly correlated topological insulator SmB$_6$ has been determined with high resolution resonant inelastic x-ray scattering (RIXS) at the Sm M$_5$ edge. The valence selectivity of RIXS allows isolating the crystal-field-split excited multiplets of the Sm$^{3+}$ (4$f^5$) configuration from those of Sm$^{2+}$ (4$f^6$) in intermediate valent SmB$_6$. The very large energy range  of RIXS allows the crystal-field analysis of a high lying multiplet at about 2.4\,eV that has the same total angular momentum $J$ as the ground state so that ambiguities due to the elastic tail can be avoided. We find that the $\Gamma_7$ doublet and $\Gamma_8$ quartet of the $^6H_{5/2}$ Hund's rule ground state are split by $\Delta^{CF}_{^6H_{5/2}}$\,=\,20$\pm$10\,meV which sets an upper limit for the 4$f$ band width. This indicates an extremely large mass renormalization from the band structure value, pointing out the need to consider the coefficients of fractional parentage for the hopping of the 4$f$ electrons.
\end{abstract}


\maketitle
SmB$_6$ is an intermediate valent Kondo insulator in which the hybridization of localized 4$f$ electrons and the conduction band ($cf$-hybridization) leads to the formation of a gap $\Delta_{h}$\,\cite{Menth1969,Cohen1970,Allen1979,Gorshunov1999,Riseborough2000} of the order of 20\,meV\,\cite{Xu2013,Zhu2013,Neupane2013,Jiang2013,Denlinger2013, Denlinger2014}. Accordingly, the resistivity increases with decreasing temperature but instead of diverging it reaches a plateau below about 10\,K.  Surface states could be an explanation for the finite low temperature conductivity\,\cite{Hatnean2013, Zhang2013,Kim2013,Wolgast2013,Kim2014,Wolgast2015,Thomas2016,Nakajima2016} and indeed it was theoretically predicted that SmB$_6$ has all the ingredients, like strong spin-orbit coupling and electrons of opposite parity ($d$ and $f$), for being a strongly correlated topological insulator\,\cite{Dzero2010,Takimoto2011,Lu2013,Alexandrov2013}. This prediction initiated many studies like angle-resolved photoelectron spectroscopy (ARPES)\,\cite{Xu2013,Zhu2013,Neupane2013,Jiang2013,Frantzeskakis2013,Denlinger2013,Denlinger2014,Xu2014,Xu2014natcomm,Hlawenka2015,Ohtsubo2019}, scanning tunneling spectroscopy\,\cite{Ruan2014,Roessler2014,Roessler2016,Jiao2016}, or de Haas-van Alphen (dHvA)\,\cite{Li2014,Tan2015,Thomas2018}. Yet, despite all these efforts, the exciting question whether these surface states are topologically non-trivial still remains to be answered. 

The $cf$-hybridization is also responsible for the intermediate valent character of Sm in SmB$_6$. At low temperatures valences of 2.5 to 2.7 have been reported\,\cite{Allen1980,Tarascon1980,Mizumaki2009,Butch2016,Utsumi2017} so that the electronic configuration of Sm is described by the Hund's rule ground states of the Sm $f^6$ (2+) and Sm $f^5$ (3+) configurations.

\begin{figure*}[t]
    \includegraphics[width=1.70\columnwidth]{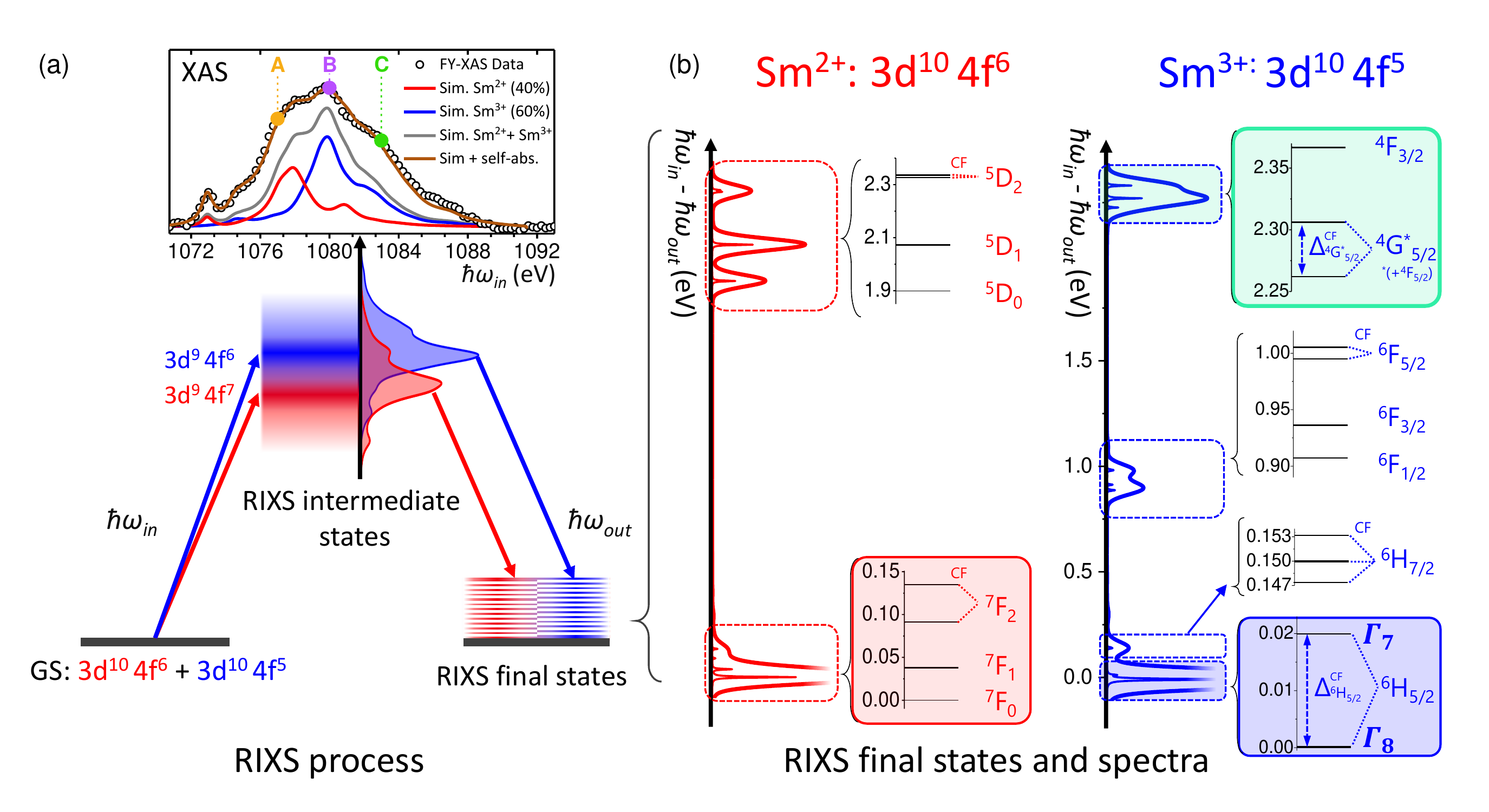}
    \caption{(color online) a) RIXS process at the Sm $M_5$-edge (3$d$\,$\rightarrow$\,4$f$) from an intermediate valent ground state of the two Sm configurations 4$f^6$ (red) and 4$f^5$ (blue) - see text. Inset: experimental, bulk sensitive fluorescence-yield x-ray absorption spectrum (FY-XAS) of Sm $M_5$-edge of SmB$_6$ (black circles), the XAS simulation (gray line) decomposed into 60\% Sm$^{3+}$ (blue line) and 40\% Sm$^{2+}$ (red line) spectral weights according to \cite{Allen1980,Tarascon1980,Mizumaki2009,Butch2016,Utsumi2017}, plus the XAS simulation including self-absorption (brown line)\,\cite{Pellegrin1993}. For graphical clarity the XAS simulations are scaled down by a factor of three. The colored dots \textsl{\textbf{ A}}, \textsl{\textbf{B}}, and \textbf{\textsl{C}} resemble the incident energies $\hbar\omega_{in}$ used in the RIXS experiment. b) Calculated RIXS spectra ($\hbar\omega_{in}$\,-\,$\hbar\omega_{out}$) of both Sm valence states (Sm$^{2+}$ red [$\hbar\omega_{in}$\,=\,\textsl{\textbf{B}}], Sm$^{3+}$ blue [$\hbar\omega_{in}$\,=\,\textsl{\textbf{C}}]) for the geometry shown in Figure\,\ref{fig3}(a) and vertical ($\sigma$) polarization of the incident photons. The red, blue, and green boxes next to the RIXS spectra show energy levels on expanded scales so that multiplet as well as expected crystal-field splittings are resolved. Thicker lines stand for higher degeneracies. The splitting $\Delta^{CF}_{^4G^*_{5/2}}$ (green box) is used for determining  $\Delta^{CF}_{^6H_{5/2}}$ (blue box).}
    \label{fig1}
\end{figure*}

\begin{figure}[t]
		\includegraphics[width=0.9\columnwidth]{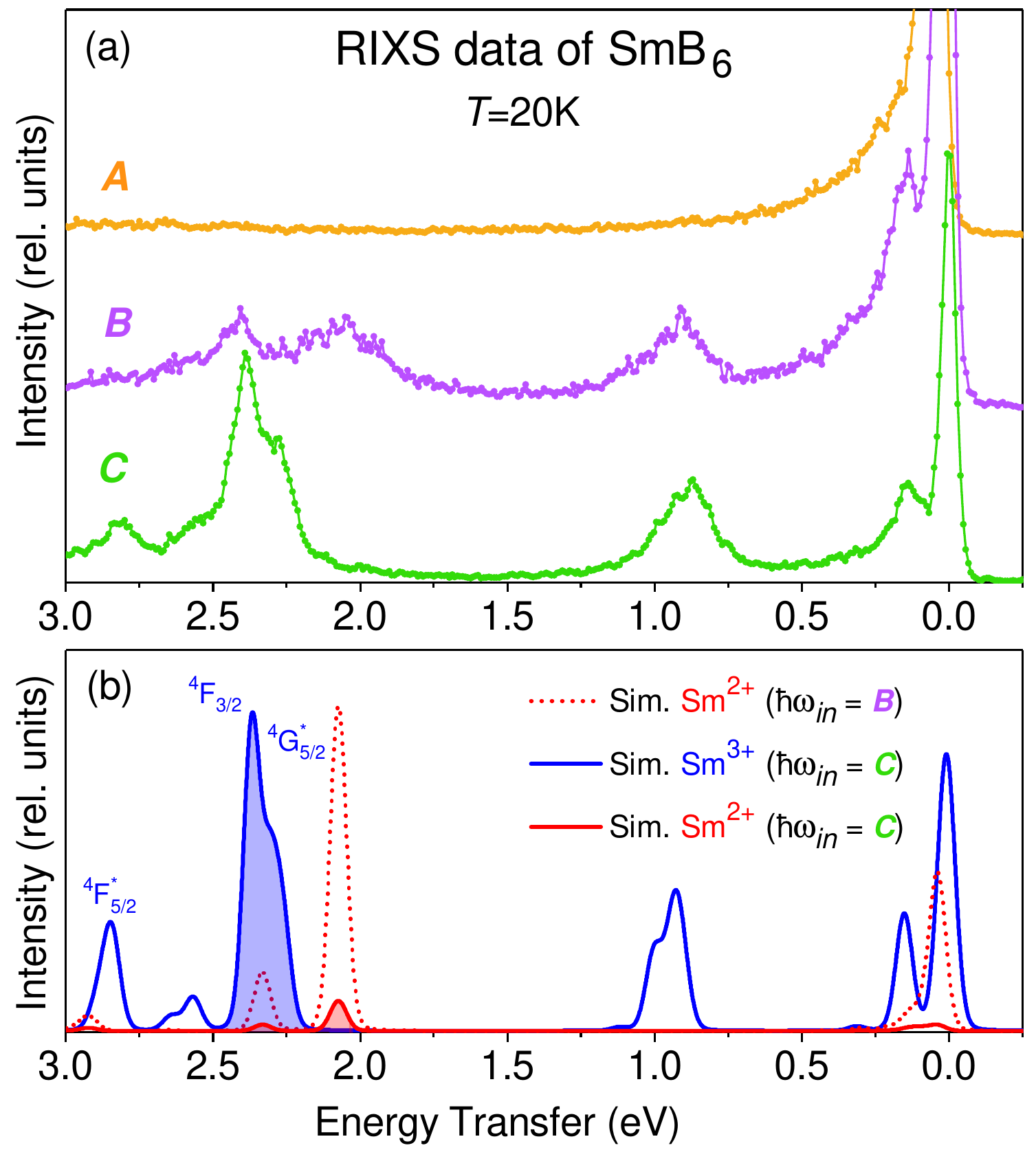}
    \caption{(color online) (a) RIXS data of SmB$_6$ for incident energies $\hbar\omega_{in}$ \textsl{\textbf{ A}}, \textsl{\textbf{B}}, and \textbf{\textsl{C}} as defined in Fig.\,1\,(a) (see text for the scattering geometry). (b) RIXS simulation for Sm$^{3+}$ (solid blue line) with $\hbar\omega_{in}$\,=\,\textsl{\textbf{C}} and for Sm$^{2+}$ with $\hbar\omega_{in}$\,=\,\textsl{\textbf{B}} (dashed red line) and \,\textsl{\textbf{C}} (solid red line) (simulation see text). 
		}
    \label{fig2}
\end{figure}

The surface topology is a bulk property so that it is indispensable to have knowledge of the parities, symmetries and near ground state energy scales of the participating bulk states. The two low lying Sm$^{2+}$\,4$f^6$  multiplets with the total angular momenta $J$\,=\,0 and $J$\,=\,1 are not CF split and their wave functions are spherical\,\cite{Lea1962}. In contrast, the cubic CF splitting of the $\Gamma_8$ quartet and $\Gamma_7$ doublet of the lowest energy multiplet $^6H_{5/2}$ of Sm$^{3+}$\,4$f^5$ has eluded its determination till today. 

Band structure calculations have been very successful in the field of semiconducting topological insulators, but they are not adequate for the rare earths because of correlations, nor are they accurate enough because the energy scales are much smaller. For example, several density functional theory calculations imply that the hole of the 4$f^5$ configuration resides in the doublet $\Gamma_7$\,\cite{Lu2013,Yanase1992,Antonov2002,Kang2015,Singh2018} but a recent hard x-ray $non$\,-\,$resonant$ inelastic x-ray scattering (NIXS) investigation\,\cite{Sundermann2018,footnote1} by some of the authors of the present study reveals that the ground-state symmetry of the Sm$^{3+}$ configuration is the $\Gamma_8$ quartet. Band structure calculations also suggest energy scales of the order of hundred meV for $\Delta^{CF}_{^6H_{5/2}}$ (see e.g.\,\cite{Kang2015,Singh2018}) although the extrapolation of the CF parameters within the REB$_6$ series suggests a splitting of the order of 15\,meV; an extrapolation that is, of course, only valid in diluted systems\,\cite{Loewenhaupt1986,Frick1986}. Along the same line, band structure calculations produce 4$f$ band widths of several hundred meV, while so far no 4$f$ dispersions in ARPES have been observed within the experimental resolution\,\cite{Xu2013,Zhu2013,Neupane2013,Jiang2013,Frantzeskakis2013,Denlinger2013,Denlinger2014,Xu2014,Xu2014natcomm,Hlawenka2015,Ohtsubo2019}. The inclusion of correlation effects using Gutzwiller or dynamical mean field approaches (DMFT) \cite{Lu2013,Kim2014a} does produce narrower bands and smaller CF splittings but it is not clear whether the mass renormalizations used or found are realistic.

Inelastic neutron scattering (INS) is the obvious technique for tackling this problem, but although providing very useful information on SmB$_6$, INS has not been successful in finding $\Delta^{CF}_{^6H_{5/2}}$. The strong neutron absorption of Sm and B even in double isotopic samples, the superposition of both Sm configurations, and the presence of $cf$-hybridization cause serious complications. Nevertheless, the following pieces of information have been obtained by INS: The spin orbit transitions $^7F_0$\,$\rightarrow$\,$^7F_1$ and $^6H_{5/2}$\,$\rightarrow$\,$^6H_{7/2}$ at $\approx$35\,meV and $\approx$130\,meV have been observed\,\cite{Alekseev1993,Alekseev1995} (see near-ground state multiplets in red and blue boxes of Fig.\,\ref{fig1}\,(b)). At low temperatures a long living spin resonance at about 14\,meV shows up in the poly- and single crystalline data\,\cite{Alekseev1993,Alekseev1995} at the $X$ and $R$ high symmetry points with a form factor that is not 4$f$-like\,\cite{Fuhrman2015}. More recent INS data show that the spin resonance with an intrinsic width of about 0.1\,meV (FWHM) decays above 30\,K and that at low temperatures no magnetic intensity is observed below the energy of this resonance which is typical for a spin gap\,\cite{Fuhrman2018}. At 100\,K i.e. well above the temperature at which the many body spin resonance disappears, quasielastic magnetic intensity ($\Gamma$/2\,$\approx$\,10\,meV HWHM) following the Sm$^{3+}$ magnetic form factor has been extracted by Alekseev \textsl{et al.} after a very careful intensity examination\,\cite{Alekseev2016}. This is suggestive of the recovery of the single ion, though broadened, magnetic spectrum, but the size the CF splitting of the Sm$^{3+}$ Hund's rule ground state remains undetermined. We note that also the high resolution ARPES studies so far have not been able to detect the CF splitting, unlike in for example YbIr$_2$Si$_2$\,\cite{Vyalikh2010}, which maybe due to the complications of the SmB$_6$ surface\,\cite{Zhu2013,Roessler2014,Roessler2016}.

Here resonant inelastic x-ray scattering (RIXS) is a promising option\,\cite{Amorese_a,Amorese_b}.  RIXS is not only element, it is also configuration selective. This is well known from studying valences at the rare earth $L$-edge in the so called partial florescence yield mode\,\cite{Kotani2001,Dallera2002}. Here we use the configuration selectivity at the $M_{4,5}$-edge (3$d$\,$\rightarrow$\,4$f$) to distinguish the excitation spectra of the two Sm configurations. 

Figure\,\ref{fig1} shows the $M_5$-edge RIXS process for SmB$_6$. The initial state configuration is an admixture of Sm$^{2+}$: 3$d^{10}$4$f^6$ (red) and Sm$^{3+}$: 3$d^{10}$4$f^5$ (blue). The resonant absorption of an $\approx$1090\,eV x-ray photon at the M$_5$ edge (3$d_{5/2}$\,$\rightarrow$\,4$f$) creates a core hole. In this intermediate state the absorption lines of the two configurations are split in energy due to the different impact of the core hole potential on either configuration. Finally, in RIXS spectroscopy the intensity of the photons emitted by the resonant radiative decay is monitored as a function of the outgoing photon energy ($\hbar\omega_{out}$) so that energy transfer spectra can be measured. In principle the decay process in the RIXS process of SmB$_6$ yields the superposition of two multiplet spectra (see simulations of two independent configurations in Fig.\,\ref{fig1}\,(b)) but the choice of the incident photon energy $\hbar\omega_{in}$ along the XAS edge allows enhancing the signal of one of the two configurations. It is possible to resolve the CF splittings in a RIXS experiment because the large life time broadening of the intermediate state does not enter, i.e. the life time broadening in RIXS that matters is that of the final state\,\cite{Haak1978,Jensen1989,Berman1991,Carra1995}.  

Figure\,\ref{fig1}\,(b) shows calculation of RIXS spectra for pure Sm$^{2+}$ (red) and pure Sm$^{3+}$ (blue). The photon-in photon-out RIXS process yields the selection rule $\Delta J$\,=\,0,\,$\pm$1,\,$\pm$2 so that multiplets with $J$\,=\,0,\,1,\,2 (for Sm$^{2+}$) and $J$\,=\,1/2,\,3/2,\,5/2,\,7/2, and 9/2 (for Sm$^{3+}$) are accessible, the latter ones being so weak that they are not shown. In the cubic point symmetry of SmB$_6$ only multiplets with $J$\,$\geq$\,2 are CF split as shown on an enlarged energy scale in the colored boxes of Fig.\,\ref{fig1}\,(b).  

Apart from the valence selectivity, another advantage of RIXS is that the transferred energy is, in contrast to INS, practically unlimited, i.e. with RIXS we can study higher lying multiplets instead of the strongly hybridized Hund's rule ground state of Sm$^{3+}$ (blue box in Fig.\,\ref{fig1}\,(b)). We will show that we can take advantage of the CF effect on the $^4G^*_{5/2}$ multiplet at about 2.4\,eV (see the green box in Fig.\,\ref{fig1}\,(b)). The asterisk indicates that due to the particularly strong intermultiplet mixing acting on this level, $L$ is no longer a good quantum number so that the multiplet labeling is not strictly valid. The total angular momentum $J$\,=\,5/2, however, remains a good quantum number for CF splittings smaller than the SO splittings. $^4G^*_{5/2}$ and the Hund's rule ground state $^6H_{5/2}$ have the same $J$ so that the same CF parameter $\check A_4^0$\,\cite{footnote2} (together with $\check A_4^4=\sqrt{5/14}\check A_4^0$) determines the CF splitting. The size of the splitting is given by $\check A_4^0$$\cdot$$\tilde{\beta}$$_{JLS}$ whereby $\tilde{\beta}$$_{JLS}$ is something like a Stevens factor that is calculated within the full multiplet routine, while $\check A_4^0$ is determined experimentally. Hence, we can gain information on the splitting of the lowest energy $^6H_{5/2}$ multiplet by fitting the RIXS signal of the $^4G^*_{5/2}$ multiplet. For $^4G^*_{5/2}$ the $\tilde{\beta}$$_{JLS}$ factor is larger than for $^6H_{5/2}$ (approximately double) so that the CF splitting is larger and less hampered by the limited energy resolution at the Sm $M_5$-edge. In addition, the signal is free of the strong tail of the elastic peak (at 0\,eV) and of the signal from other low energy excitations.

The SmB$_6$\,\cite{Kim2014} $M$-edge RIXS experiment at 20\,K was performed at the ERIXS spectrometer of the ID32 beamline\,\cite{Brookes2018} at the European Synchrotron Radiation Facility (ERSF), Grenoble, France with a resolution of 45\,meV at the Sm $M_5$-edge ($\approx$\,1090\,meV). Data were taken with two different scattering angles, namely 2$\Theta$\,=\,90$^{\circ}$ and 150$^{\circ}$. Further details of the set-up are given in the Appendix. Simulation were performed with the full multiplet code \textsl{Quanty}\,\cite{Haverkort2012,Haverkort2016}. Atomic parameters were taken from the Cowan code\,\cite{CowanBook} and the reduction factor of the Slater integrals $r_{4f-4f}$\,=$r_{3d-4f}$\,=\,0.86 were used (see Appendix). These values are in agreement with those in Ref.\,\cite{Sundermann2018}.

The inset of Fig\,\ref{fig1}\,(a) shows the bulk-sensitive experimental fluorescence-yield XAS (FY-XAS) data of the Sm $M_5$-edge of SmB$_6$ at 20\,K, with the photon polarization parallel to the 100 direction (black line). These data have been simulated by calculating an XAS spectrum (gray line) containing Sm$^{3+}$ (40\% ) and Sm$^{2+}$ (60\%) spectral weights according to the SmB$_6$ valence at low $T$\,\cite{Allen1980,Tarascon1980,Mizumaki2009,Butch2016,Utsumi2017}. Then self-absorption effects were included in the simulation (see brown line)\,\cite{Pellegrin1993} and compared with the FY-XAS data. Note, for reasons of graphical clarity the XAS data have been rescaled by a factor of three. The orange, purple and green dots marked \textbf{\textsl{A}}, \textbf{\textsl{\textsl{B}}} and \textbf{\textsl{\textsl{C}}} indicate the incident energies that were used for the RIXS experiment.   

Figure\,\ref{fig2}\,(a) shows the RIXS data at $T$\,=\,20\,K up to 3\,eV taken with the three different incident energies $\hbar\omega_{in}$\,=\,\textbf{\textsl{A}},\textbf{\textsl{B}}, and \textbf{\textsl{C}} and a scattering angle of $2\Theta$\,=\,90$^{\circ}$. $\hbar\omega_{in}$\,=\,\textbf{\textsl{A}} corresponds to the pre-edge region where the $3d$\,$\rightarrow$\,$4f$ absorption process is dominated by the ground state of Sm$^{2+}$ $4f^6$. The asymmetric intensity close to the elastic line is indicative for the low energy transitions $^7F_0$\,$\rightarrow$\,$^7F_1$ at 35\,meV and some $^7F_0$\,$\rightarrow$\,$^7F_2$ at about 150\,meV, whereas higher energy transfers have no cross-section because they require larger incident energies due to selection rules. At $\hbar\omega_{in}$\,=\,\textbf{\textsl{C}} the absorption arises mainly from the $4f^5$ ground state of Sm$^{3+}$. Energy \textbf{\textsl{B}} is in-between, i.e. the RIXS spectrum shows features characteristic of both valences but is not simply the superposition of spectrum \textbf{\textsl{A}} and \textbf{\textsl{C}} because of the incident energy dependence of the accessible excitations. Figure\,\ref{fig2}\,(b), shows full multiplet RIXS calculations for the same spectrometer configuration for Sm$^{2+}$ with the incident energies \textbf{\textsl{B}} (dotted red line) and \textbf{\textsl{C}} (red solid line) and and for Sm$^{3+}$ with incident energy \textbf{\textsl{C}} (solid blue line) . The comparison of both panels demonstrates the energy selectivity of the RIXS signal and it confirms that the spectrum measured with $\hbar\omega_{in}$\,=\,\textbf{\textsl{C}} resembles almost purely Sm$^{3+}$ multiplets. We will therefore focus on the region of the $^4F_{3/2}$ and $^4G^*_{5/2}$ multiplets measured with this incident energy for further analysis of the crystal-field problem of Sm$^{3+}$ (see colored regions in Fig.\,\ref{fig2}\,(b). 

\begin{figure}[t]
    \includegraphics[width=0.9\columnwidth]{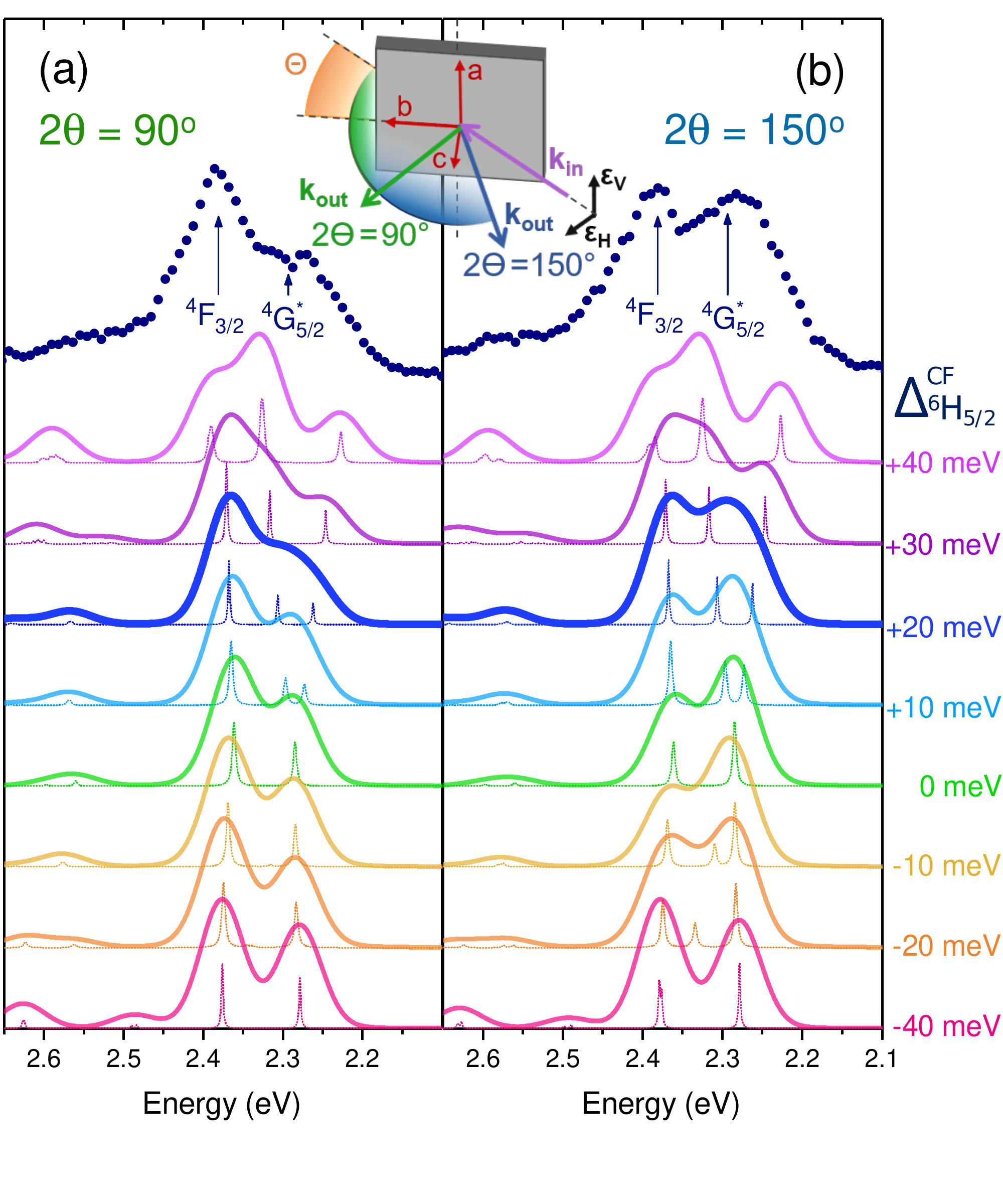}
    \caption{(color online) Data and simulations of RIXS spectra with $\hbar\omega_{in}$\,=\,\textbf{\textsl{C}} (dark blue dots) with horizontal polarization $\epsilon_H$ ($\pi$) for two different scattering angles, (a) for 2$\Theta$\,=\,90$^{\circ}$ and (b) for 2$\Theta$\,=\,150$^{\circ}$ (see inset). Different colors are simulations with different crystal-field splittings; thin dotted lines with an unrealistic narrow resolution, solid thick lines taking into account the 45 meV Gaussian resolution function. The numbers refer to the respective Hund's rule ground state splitting $\Delta^{CF}_{^6H_{5/2}}$, positive numbers refer to a $\Gamma_8$ and negative numbers to a $\Gamma_7$ ground state. Note, $\Delta^{CF}_{^4G^*_{5/2}}$\,$\approx$\,2.2$\Delta^{CF}_{^6H_{5/2}}$.}
    \label{fig3}
\end{figure}

The top of Fig.\,\ref{fig3}\,(a) and (b) show the RIXS data of the $^4F_{3/2}$ and $^4G^*_{5/2}$ multiplets (dark blue dots) at around 2.4\,eV energy transfer ($\hbar\omega_{in}\,=\,$\textsl{\textbf{C}}) measured with horizontal ($\pi$) polarization and two different scattering angles, 2$\theta$\,=\,90$^{\circ}$ and 150$^{\circ}$, thus taking advantage of the cross-section dependence on the scattering geometry. We recall that the multiplet $^4F_{3/2}$ is not affected by the CF because of $J$\,$<$\,2 but that a finite CF splits the $^4G^*_{5/2}$ multiplet into two levels.

Figure\,\ref{fig3} also shows simulations broadened with a 45\,meV Gaussian resolution function for different CF splittings. We find that, for the same CF parameter, $\Delta^{CF}_{^4G^*_{5/2}}$ is about 2.2 times larger than $\Delta^{CF}_{^6H_{5/2}}$.  We show the simulations for $\Delta^{CF}_{^6H_{5/2}}$\,=\,$\left[-40,+40\right]$\,meV in steps of 10\,meV, whereby the positive numbers refer to a $\Gamma_8$ and the negative ones to a $\Gamma_7$ ground state.  Here the narrow thin lines correspond to the same CF simulation but with an unrealistic small resolution in order to visualize the details of the CF splittings.  The simulation with $\Delta^{CF}_{^6H_{5/2}}$\,=\,0 (green lines) shows two main peaks, the $^4G^*_{5/2}$ multiplet and the $^4F_{3/2}$ about 80\,meV higher in energy.  We learn from these simulations that for CF splittings of less than 40\,meV the intermixing of the two multiplets is negligible. We now compare in detail data and simulations: For 0\,meV CF (green lines) and for $+$10 and $-$10\,meV splitting  (light blue and yellow lines) and 2$\Theta$\,=\,150$^{\circ}$ the $^4G^*_{5/2}$ intensity would be stronger than the $^4$F$_{3/2}$ peak, see Fig.\,\ref{fig3}\,(b). This is not the case in the experiment. Hence, the CF splitting of the ground state must be larger that 10\,meV. For a negative CF splitting only one $^4G^*_{5/2}$ CF excitation would have intensity in the 2$\Theta$\,=90$^{\circ}$ configuration, thus leading to a deep valley between the two multiplets which has not been observed. see Fig.\,\ref{fig3}\,(a). We therefore conclude that the splitting must be positive, i.e. we confirm the results of previous directional dependent NIXS data\,\cite{Sundermann2018}. For $+$40\,meV CF splitting the spectral shape has changed considerably for both scattering geometries so we also exclude this possibility as well. It turns out that peak shapes and intensity ratios of both scattering configurations are best reproduced with $\Delta^{CF}_{^6H_{5/2}}$\,=\,+20\,meV. 

\begin{figure}[t]
    \includegraphics[width=0.9\columnwidth]{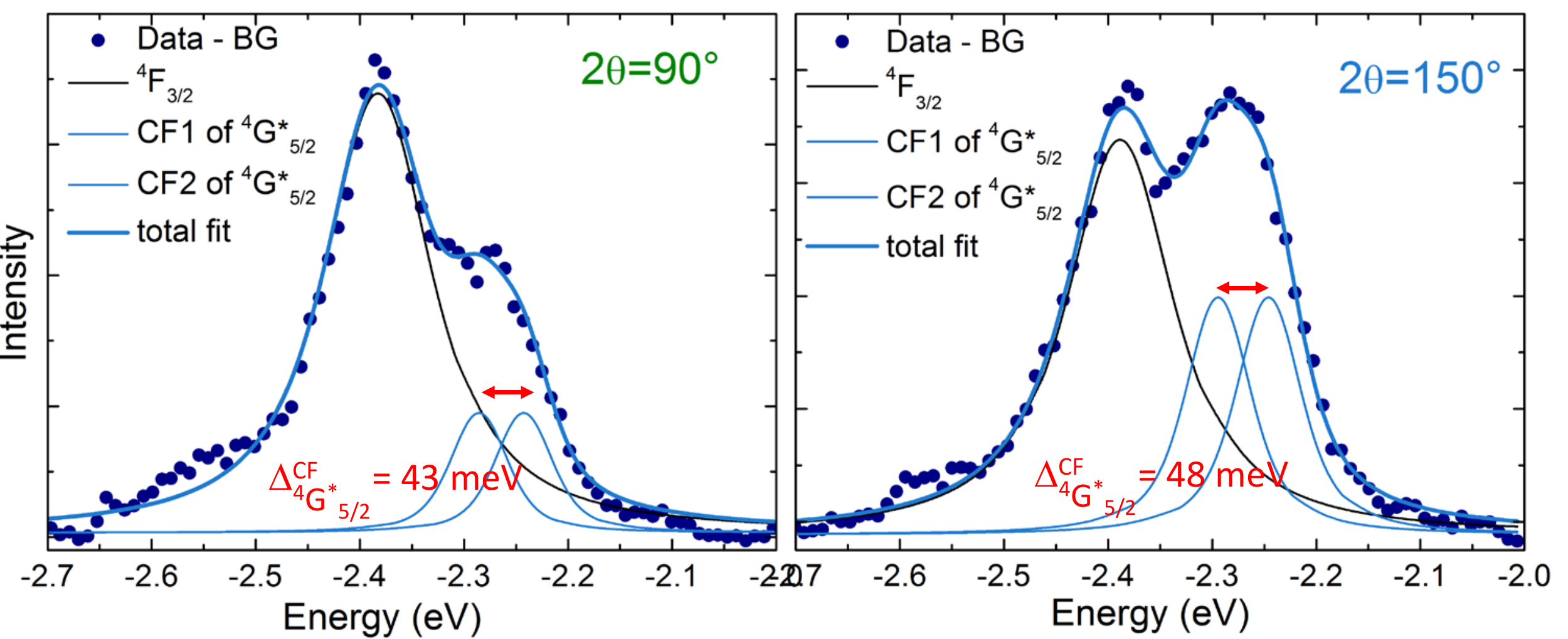}
    \caption{(color online) Background corrected RIXS spectra of Fig.\,\ref{fig3} with an empirical fit of three Voigt profiles (see text), one resembling the $^4F_{3/2}$ multiplet and two the crystal-field split $^4G^*_{5/2}$. The crystal-field splitting of the $^4G^*_{5/2}$ multiplets corresponds to $\Delta^{CF}_{^6H_{5/2}}$\,=\,20 and 22\,meV, respectively.}
    \label{fig4}
\end{figure}

Figure\,\ref{fig4} shows the same RIXS data as in Fig.\,\ref{fig3} but after subtracting a linear background. The lines represent an empirical fit with three Voigt profiles whereby the Gaussian contribution is kept fixed to the experimental resolution. The Lorentzian widths, the line positions and intensities were varied with the simplification that lifetime broadening and intensity of the two CF excitations are identical. The best fits yield $\Delta^{CF}_{^4G^*_{5/2}}$\,=\,43 and 48\,meV  for the 2$\Theta$\,=\,90$^{\circ}$ and 150$^{\circ}$ scattering configuration, respectively, corresponding to a splitting of 20 and 22\,meV of the ground state multiplet $^6H_{5/2}$. Other trials with larger crystal-field splittings no longer reproduce the data, see Appendix. We learn form this exercise that $\Delta^{CF}_{^4G^*_{5/2}}$ ($\Delta^{CF}_{^6H_{5/2}}$) should be $<$66\,meV ($<$30\,meV). Summarizing, we thus find $\Delta^{CF}_{^6H_{5/2}}$\,=\,20$\pm$10\,meV.

The present RIXS result agrees surprisingly well with the CF splitting that is expected from the extrapolation within the REB$_6$ series\,\cite{Loewenhaupt1986}. The result also explains the lineshape of the lowest $f$ state signal as measured in photoemission\,\cite{Denlinger2014}; we can now propose to describe it in terms of two Lorenzian lines, one twice as strong as the other according to a $\Gamma_8$ quartet ground state and a $\Gamma_7$ excited doublet, that are about 20\,meV apart. Furthermore, the present data confirm the $non$\,-\,$resonant$ inelastic x-ray scattering result of SmB$_6$ that also finds a quartet ground state\,\cite{Sundermann2018,footnote1}.

The finding that the CF splitting is 10\,meV\,$<$\,$\Delta^{CF}_{^6H_{5/2}}$\,$<$\,30\,meV in combination with the NIXS result that the ground state is not a highly mixed  $\Gamma_8$ and $\Gamma_7$ state\,\cite{Sundermann2018} indicates that the 4$f$ band width is small and less than 30\,meV. Considering the fact that the band width from band structure calculations is several hundred meV, we infer that the mass renormalization is extremely large. This also gives credit to the idea that coefficients of fractional parentage should be considered for removing or adding an electron from/to the lowest Sm $f^6$ or $f^5$ multiplet states\,\cite{Sawatzky}: a reduction factor of 0.033 can be found for the $f$\,--\,$f$ hopping. A Gutzwiller study uses a somewhat less strong reduction factor\,\cite{Lu2013}, while a DMFT calculation\,\cite{Kim2014a} found indeed the extremely narrow bands. It should be noted however, that the sign of the CF splitting and thus also its magnitude used or found in these many body calculations,\cite{Lu2013,Kim2014a} is different from the experiment. It would be highly desirable if these calculations could be tuned in a way that they reproduce the experimental values.

The electronic configuration selectivity, the large accessible energy transfers, and the cross-section dependence on the scattering configuration in RIXS have been instrumental to observe finger prints of the CF splitting of the Sm$^{3+}$ Hund's rule ground state in SmB$_6$. 
We find $\Delta^{CF}_{^6H_{5/2}}$\,=\,20$\pm$10\,meV describes the data well, thereby setting limits to the 4$f$ band width.

\section{Appendix}
\subsection{Sample and Experiment}
The RIXS experiment was performed on aluminum flux grown single crystals\,\cite{Kim2014} that were aligned by Laue prior to the experiment. The data were cleaved $in$ $situ$ under vacuum, then transferred to the main chamber and measured at 20\,K. Data were acquired for about 5\,hours for each spectrum (only 3\,hours for the spectrum \textsl{\textbf{B}}). The instrument 45\,meV-FWHM Gaussian response function was estimated by measuring a carbon tape. The measurements were performed with horizontal polarization ($\pi$) of the incident photons, two different scattering angles, $2\Theta$\,=\,90$^{\circ}$ and $2\Theta$\,=\,150$^{\circ}$, a sample angle of $\theta=37.3^\circ$ and with the \textbf{\textsl{b}} and \textbf{\textsl{c}} directions of the sample in the scattering plane (see inset of Fig. 3).

\begin{figure}[]
		\includegraphics[width=0.9\columnwidth]{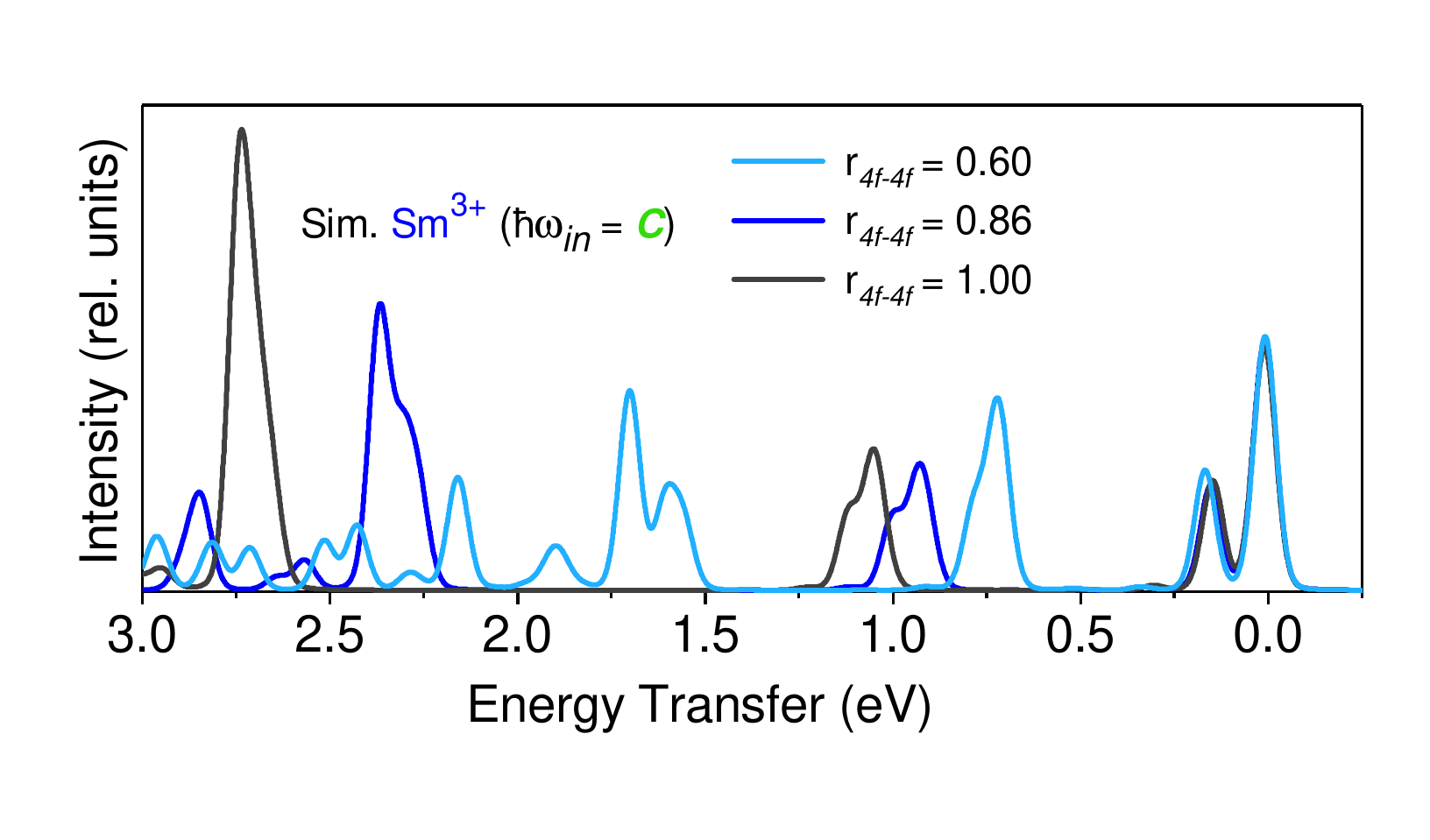}
	    \caption{(color online) RIXS simulations of Sm$^{3+}$ for incident energy \textbf{\textsl{C}} with various reduction factors $r_{4f-4f}$ ($r_{3d-4f}=0.86$).}
    \label{}
\end{figure}

\begin{figure}[]
    \includegraphics[width=0.9\columnwidth]{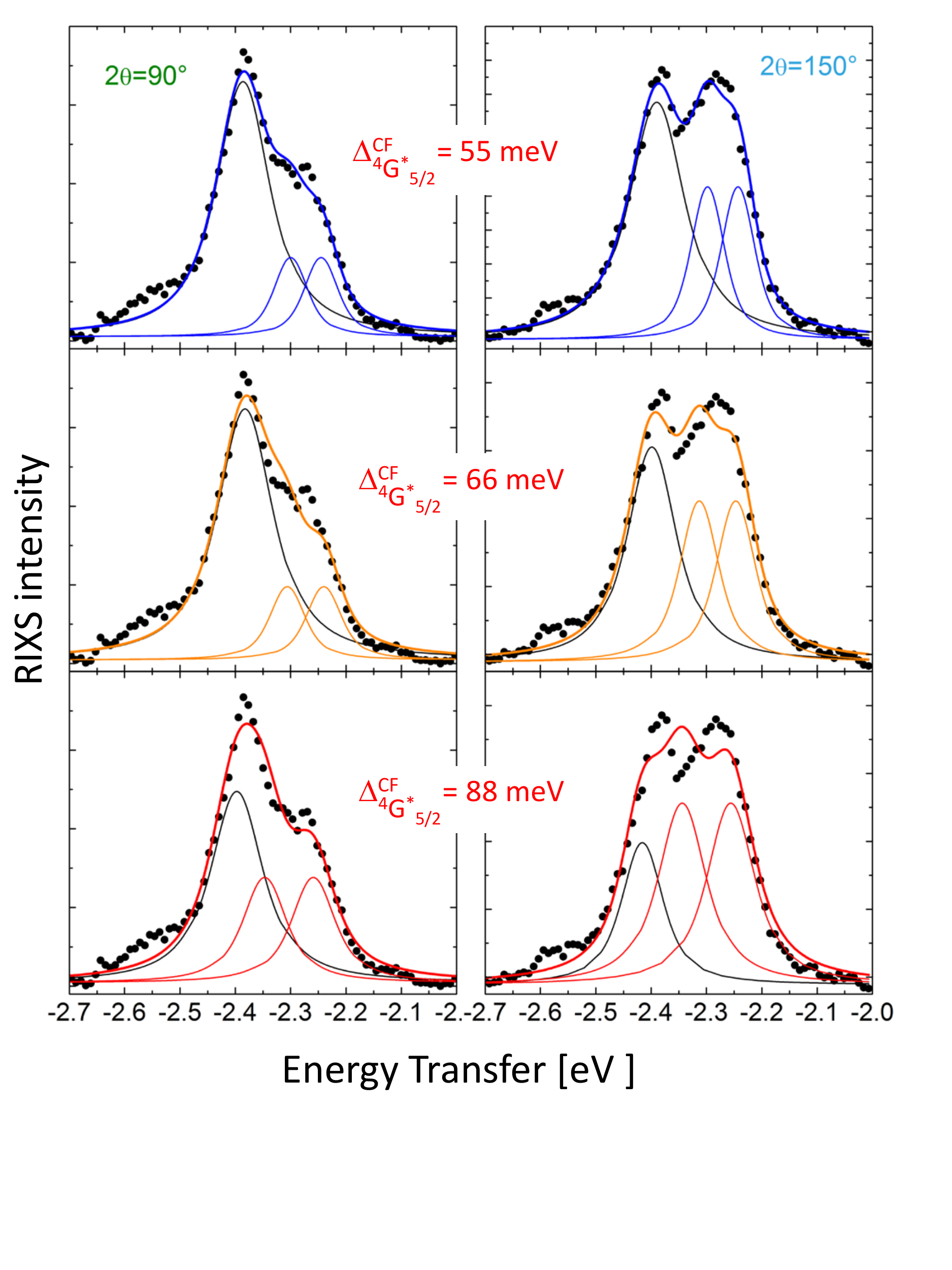}
	    \caption{(color online)  Simulations with three Voigt profiles of the $^4$F$_{3/2}$ and crystal-field split $^4G^*_{5/2}$ multiplets measured with 2$\theta$\,=\,90$^{\circ}$ and 150$^{\circ}$. The simulations assume different crystal-field splittings. The respective splittings are given in the panels for $^4G^*_{5/2}$. The splitting of the ground state multiplet $^6H_{5/2}$ is 2.2 times smaller.}
	\label{}
\end{figure}

\subsection{Simulation}
Simulations were performed with the full multiplet code \textsl{Quanty}\,\cite{Haverkort2012,Haverkort2016}. Atomic parameters were taken from the Cowan code\,\cite{CowanBook}. Figure\,5 shows that the energy positions of the multiplets depend are very sensitive to changes of the reduction factor $r_{4f-4f}$ of the Slater integrals so that $r_{4f-4f}$ was determined by adjusting the energy positions of the multiplet excitations. The reduction factor $r_{3d-4f}$, on the other hand, affects only slightly the relative intensities of the RIXS peaks. We find that $r_{4f-4f}$\,=\,$r_{3d-4f}$\,=\,0.86 provides a very good fit of the relative RIXS intensities and line positions, and to the XAS data. These values are in agreement with those in Ref.\,\cite{Sundermann2018}.

Figure\,6 shows empirical descriptions of the $^4F_{3/2}$ and $^4G^*_{5/2}$ multiplets with three Voigt profiles, one for the $^4F_{3/2}$ and two for the crystal-field split $^4G^*_{5/2}$ multiplet. The Gaussian contribution was kept fixed to the instrumental resolution of 45\,meV, while the Lorentzian line widths were varied. Here the constraint was imposed that the two crystal-field excitations have the same line width and also the same intensity. The position of the three lines was varied with the limitation that the separation of the crystal-field excitations was set to specific values (see panels of Fig.\,6). For $\Delta^{CF}_{^4G^*_{5/2}}$\,=\,55\,meV ($\Delta^{CF}_{^6H_{5/2}}$\,=\,25\,meV) both configurations are still well described with the three Voigt profiles, for $\Delta^{CF}_{^4G^*_{5/2}}$\,$\geq$\,66\,meV ($\Delta^{CF}_{^6H_{5/2}}$\,$\geq$\,30\,meV) it is no longer possible to describe the data with the scattering angle of 2$\theta$\,=\,150$^{\circ}$. This shows that the crystal-field splitting of the ground state must be smaller than 30\,meV.

\section{Acknowledgment} A.A. and A.S. gratefully acknowledge the financial support of the Deutsche Forschungsgemeinschaft under project SE\,1441-4-1.


\begin{thebibliography}{72}%
\makeatletter
\providecommand \@ifxundefined [1]{%
 \@ifx{#1\undefined}
}%
\providecommand \@ifnum [1]{%
 \ifnum #1\expandafter \@firstoftwo
 \else \expandafter \@secondoftwo
 \fi
}%
\providecommand \@ifx [1]{%
 \ifx #1\expandafter \@firstoftwo
 \else \expandafter \@secondoftwo
 \fi
}%
\providecommand \natexlab [1]{#1}%
\providecommand \enquote  [1]{``#1''}%
\providecommand \bibnamefont  [1]{#1}%
\providecommand \bibfnamefont [1]{#1}%
\providecommand \citenamefont [1]{#1}%
\providecommand \href@noop [0]{\@secondoftwo}%
\providecommand \href [0]{\begingroup \@sanitize@url \@href}%
\providecommand \@href[1]{\@@startlink{#1}\@@href}%
\providecommand \@@href[1]{\endgroup#1\@@endlink}%
\providecommand \@sanitize@url [0]{\catcode `\\12\catcode `\$12\catcode
  `\&12\catcode `\#12\catcode `\^12\catcode `\_12\catcode `\%12\relax}%
\providecommand \@@startlink[1]{}%
\providecommand \@@endlink[0]{}%
\providecommand \url  [0]{\begingroup\@sanitize@url \@url }%
\providecommand \@url [1]{\endgroup\@href {#1}{\urlprefix }}%
\providecommand \urlprefix  [0]{URL }%
\providecommand \Eprint [0]{\href }%
\providecommand \doibase [0]{http://dx.doi.org/}%
\providecommand \selectlanguage [0]{\@gobble}%
\providecommand \bibinfo  [0]{\@secondoftwo}%
\providecommand \bibfield  [0]{\@secondoftwo}%
\providecommand \translation [1]{[#1]}%
\providecommand \BibitemOpen [0]{}%
\providecommand \bibitemStop [0]{}%
\providecommand \bibitemNoStop [0]{.\EOS\space}%
\providecommand \EOS [0]{\spacefactor3000\relax}%
\providecommand \BibitemShut  [1]{\csname bibitem#1\endcsname}%
\let\auto@bib@innerbib\@empty
\bibitem [{\citenamefont {Menth}\ \emph {et~al.}(1969)\citenamefont {Menth},
  \citenamefont {Buehler},\ and\ \citenamefont {Geballe}}]{Menth1969}%
  \BibitemOpen
  \bibfield  {author} {\bibinfo {author} {\bibfnamefont {A.}~\bibnamefont
  {Menth}}, \bibinfo {author} {\bibfnamefont {E.}~\bibnamefont {Buehler}}, \
  and\ \bibinfo {author} {\bibfnamefont {T.~H.}\ \bibnamefont {Geballe}},\
  }\bibfield  {title} {\enquote {\bibinfo {title} {Magnetic and semiconducting
  properties of {SmB$_6$}},}\ }\href {\doibase 10.1103/PhysRevLett.22.295}
  {\bibfield  {journal} {\bibinfo  {journal} {Phys. Rev. Lett.}\ }\textbf
  {\bibinfo {volume} {22}},\ \bibinfo {pages} {295--297} (\bibinfo {year}
  {1969})}\BibitemShut {NoStop}%
\bibitem [{\citenamefont {Cohen}\ \emph {et~al.}(1970)\citenamefont {Cohen},
  \citenamefont {Eibsch\"utz},\ and\ \citenamefont {West}}]{Cohen1970}%
  \BibitemOpen
  \bibfield  {author} {\bibinfo {author} {\bibfnamefont {R.~L.}\ \bibnamefont
  {Cohen}}, \bibinfo {author} {\bibfnamefont {M.}~\bibnamefont {Eibsch\"utz}},
  \ and\ \bibinfo {author} {\bibfnamefont {K.~W.}\ \bibnamefont {West}},\
  }\bibfield  {title} {\enquote {\bibinfo {title} {Electronic and magnetic
  structure of {S}m{B}$_6$},}\ }\href {\doibase 10.1103/PhysRevLett.24.383}
  {\bibfield  {journal} {\bibinfo  {journal} {Phys. Rev. Lett.}\ }\textbf
  {\bibinfo {volume} {24}},\ \bibinfo {pages} {383--386} (\bibinfo {year}
  {1970})}\BibitemShut {NoStop}%
\bibitem [{\citenamefont {Allen}\ \emph {et~al.}(1979)\citenamefont {Allen},
  \citenamefont {Batlogg},\ and\ \citenamefont {Wachter}}]{Allen1979}%
  \BibitemOpen
  \bibfield  {author} {\bibinfo {author} {\bibfnamefont {J.~W.}\ \bibnamefont
  {Allen}}, \bibinfo {author} {\bibfnamefont {B.}~\bibnamefont {Batlogg}}, \
  and\ \bibinfo {author} {\bibfnamefont {P.}~\bibnamefont {Wachter}},\
  }\bibfield  {title} {\enquote {\bibinfo {title} {Large low-temperature {H}all
  effect and resistivity in mixed-valent {S}m{B}$_6$},}\ }\href {\doibase
  10.1103/PhysRevB.20.4807} {\bibfield  {journal} {\bibinfo  {journal} {Phys.
  Rev. B}\ }\textbf {\bibinfo {volume} {20}},\ \bibinfo {pages} {4807--4813}
  (\bibinfo {year} {1979})}\BibitemShut {NoStop}%
\bibitem [{\citenamefont {Gorshunov}\ \emph {et~al.}(1999)\citenamefont
  {Gorshunov}, \citenamefont {Sluchanko}, \citenamefont {Volkov}, \citenamefont
  {Dressel}, \citenamefont {Knebel}, \citenamefont {Loidl},\ and\ \citenamefont
  {Kunii}}]{Gorshunov1999}%
  \BibitemOpen
  \bibfield  {author} {\bibinfo {author} {\bibfnamefont {B.}~\bibnamefont
  {Gorshunov}}, \bibinfo {author} {\bibfnamefont {N.}~\bibnamefont
  {Sluchanko}}, \bibinfo {author} {\bibfnamefont {A.}~\bibnamefont {Volkov}},
  \bibinfo {author} {\bibfnamefont {M.}~\bibnamefont {Dressel}}, \bibinfo
  {author} {\bibfnamefont {G.}~\bibnamefont {Knebel}}, \bibinfo {author}
  {\bibfnamefont {A.}~\bibnamefont {Loidl}}, \ and\ \bibinfo {author}
  {\bibfnamefont {S.}~\bibnamefont {Kunii}},\ }\bibfield  {title} {\enquote
  {\bibinfo {title} {Low-energy electrodynamics of {S}m{B}$_6$},}\ }\href
  {\doibase 10.1103/PhysRevB.59.1808} {\bibfield  {journal} {\bibinfo
  {journal} {Phys. Rev. B}\ }\textbf {\bibinfo {volume} {59}},\ \bibinfo
  {pages} {1808--1814} (\bibinfo {year} {1999})}\BibitemShut {NoStop}%
\bibitem [{\citenamefont {Riseborough}(2000)}]{Riseborough2000}%
  \BibitemOpen
  \bibfield  {author} {\bibinfo {author} {\bibfnamefont {P.~S.}\ \bibnamefont
  {Riseborough}},\ }\bibfield  {title} {\enquote {\bibinfo {title} {Heavy
  fermion semiconductors},}\ }\href {\doibase 10.1080/000187300243345}
  {\bibfield  {journal} {\bibinfo  {journal} {Adv. Phys.}\ }\textbf {\bibinfo
  {volume} {49}},\ \bibinfo {pages} {257--320} (\bibinfo {year}
  {2000})}\BibitemShut {NoStop}%
\bibitem [{\citenamefont {Xu}\ \emph {et~al.}(2013)\citenamefont {Xu},
  \citenamefont {Shi}, \citenamefont {Biswas}, \citenamefont {Matt},
  \citenamefont {Dhaka}, \citenamefont {Huang}, \citenamefont {Plumb},
  \citenamefont {Radovi\ifmmode~\acute{c}\else \'{c}\fi{}}, \citenamefont
  {Dil}, \citenamefont {Pomjakushina}, \citenamefont {Conder}, \citenamefont
  {Amato}, \citenamefont {Salman}, \citenamefont {Paul}, \citenamefont {Mesot},
  \citenamefont {Ding},\ and\ \citenamefont {Shi}}]{Xu2013}%
  \BibitemOpen
  \bibfield  {author} {\bibinfo {author} {\bibfnamefont {N.}~\bibnamefont
  {Xu}}, \bibinfo {author} {\bibfnamefont {X.}~\bibnamefont {Shi}}, \bibinfo
  {author} {\bibfnamefont {P.~K.}\ \bibnamefont {Biswas}}, \bibinfo {author}
  {\bibfnamefont {C.~E.}\ \bibnamefont {Matt}}, \bibinfo {author}
  {\bibfnamefont {R.~S.}\ \bibnamefont {Dhaka}}, \bibinfo {author}
  {\bibfnamefont {Y.}~\bibnamefont {Huang}}, \bibinfo {author} {\bibfnamefont
  {N.~C.}\ \bibnamefont {Plumb}}, \bibinfo {author} {\bibfnamefont
  {M.}~\bibnamefont {Radovi\ifmmode~\acute{c}\else \'{c}\fi{}}}, \bibinfo
  {author} {\bibfnamefont {J.~H.}\ \bibnamefont {Dil}}, \bibinfo {author}
  {\bibfnamefont {E.}~\bibnamefont {Pomjakushina}}, \bibinfo {author}
  {\bibfnamefont {K.}~\bibnamefont {Conder}}, \bibinfo {author} {\bibfnamefont
  {A.}~\bibnamefont {Amato}}, \bibinfo {author} {\bibfnamefont
  {Z.}~\bibnamefont {Salman}}, \bibinfo {author} {\bibfnamefont {D.~McK.}\
  \bibnamefont {Paul}}, \bibinfo {author} {\bibfnamefont {J.}~\bibnamefont
  {Mesot}}, \bibinfo {author} {\bibfnamefont {H.}~\bibnamefont {Ding}}, \ and\
  \bibinfo {author} {\bibfnamefont {M.}~\bibnamefont {Shi}},\ }\bibfield
  {title} {\enquote {\bibinfo {title} {Surface and bulk electronic structure of
  the strongly correlated system {SmB}$_6$ and implications for a topological
  {K}ondo insulator},}\ }\href {\doibase 10.1103/PhysRevB.88.121102} {\bibfield
   {journal} {\bibinfo  {journal} {Phys. Rev. B}\ }\textbf {\bibinfo {volume}
  {88}},\ \bibinfo {pages} {121102} (\bibinfo {year} {2013})}\BibitemShut
  {NoStop}%
\bibitem [{\citenamefont {Zhu}\ \emph {et~al.}(2013)\citenamefont {Zhu},
  \citenamefont {Nicolaou}, \citenamefont {Levy}, \citenamefont {Butch},
  \citenamefont {Syers}, \citenamefont {Wang}, \citenamefont {Paglione},
  \citenamefont {Sawatzky}, \citenamefont {Elfimov},\ and\ \citenamefont
  {Damascelli}}]{Zhu2013}%
  \BibitemOpen
  \bibfield  {author} {\bibinfo {author} {\bibfnamefont {Z.-H.}\ \bibnamefont
  {Zhu}}, \bibinfo {author} {\bibfnamefont {A.}~\bibnamefont {Nicolaou}},
  \bibinfo {author} {\bibfnamefont {G.}~\bibnamefont {Levy}}, \bibinfo {author}
  {\bibfnamefont {N.~P.}\ \bibnamefont {Butch}}, \bibinfo {author}
  {\bibfnamefont {P.}~\bibnamefont {Syers}}, \bibinfo {author} {\bibfnamefont
  {X.~F.}\ \bibnamefont {Wang}}, \bibinfo {author} {\bibfnamefont
  {J.}~\bibnamefont {Paglione}}, \bibinfo {author} {\bibfnamefont {G.~A.}\
  \bibnamefont {Sawatzky}}, \bibinfo {author} {\bibfnamefont {I.~S.}\
  \bibnamefont {Elfimov}}, \ and\ \bibinfo {author} {\bibfnamefont
  {A.}~\bibnamefont {Damascelli}},\ }\bibfield  {title} {\enquote {\bibinfo
  {title} {Polarity-driven surface metallicity in {SmB}$_6$},}\ }\href
  {\doibase 10.1103/PhysRevLett.111.216402} {\bibfield  {journal} {\bibinfo
  {journal} {Phys. Rev. Lett.}\ }\textbf {\bibinfo {volume} {111}},\ \bibinfo
  {pages} {216402} (\bibinfo {year} {2013})}\BibitemShut {NoStop}%
\bibitem [{\citenamefont {Neupane}\ \emph {et~al.}(2013)\citenamefont
  {Neupane}, \citenamefont {Alidoust}, \citenamefont {Xu}, \citenamefont
  {Kondo}, \citenamefont {Ishida}, \citenamefont {Kim}, \citenamefont {Liu},
  \citenamefont {Belopolski}, \citenamefont {Jo}, \citenamefont {Chang},
  \citenamefont {Jeng}, \citenamefont {Durakiewicz}, \citenamefont {Balicas},
  \citenamefont {Lin}, \citenamefont {Bansil}, \citenamefont {Shin},
  \citenamefont {Fisk},\ and\ \citenamefont {Hasan}}]{Neupane2013}%
  \BibitemOpen
  \bibfield  {author} {\bibinfo {author} {\bibfnamefont {M.}~\bibnamefont
  {Neupane}}, \bibinfo {author} {\bibfnamefont {N.}~\bibnamefont {Alidoust}},
  \bibinfo {author} {\bibfnamefont {S-Y.}\ \bibnamefont {Xu}}, \bibinfo
  {author} {\bibfnamefont {T.}~\bibnamefont {Kondo}}, \bibinfo {author}
  {\bibfnamefont {Y.}~\bibnamefont {Ishida}}, \bibinfo {author} {\bibfnamefont
  {D.~J.}\ \bibnamefont {Kim}}, \bibinfo {author} {\bibfnamefont {Chang}\
  \bibnamefont {Liu}}, \bibinfo {author} {\bibfnamefont {I.}~\bibnamefont
  {Belopolski}}, \bibinfo {author} {\bibfnamefont {Y.~J.}\ \bibnamefont {Jo}},
  \bibinfo {author} {\bibfnamefont {T-R.}\ \bibnamefont {Chang}}, \bibinfo
  {author} {\bibfnamefont {H-T.}\ \bibnamefont {Jeng}}, \bibinfo {author}
  {\bibfnamefont {T.}~\bibnamefont {Durakiewicz}}, \bibinfo {author}
  {\bibfnamefont {L.}~\bibnamefont {Balicas}}, \bibinfo {author} {\bibfnamefont
  {H.}~\bibnamefont {Lin}}, \bibinfo {author} {\bibfnamefont {A.}~\bibnamefont
  {Bansil}}, \bibinfo {author} {\bibfnamefont {S.}~\bibnamefont {Shin}},
  \bibinfo {author} {\bibfnamefont {Z.}~\bibnamefont {Fisk}}, \ and\ \bibinfo
  {author} {\bibfnamefont {M.}~\bibnamefont {Hasan}},\ }\bibfield  {title}
  {\enquote {\bibinfo {title} {Surface electronic structure of the topological
  kondo-insulator candidate correlated electron system {SmB}$_6$},}\ }\href
  {\doibase 10.1038/ncomms3991} {\bibfield  {journal} {\bibinfo  {journal}
  {Nat. Commun.}\ }\textbf {\bibinfo {volume} {4}},\ \bibinfo {pages} {2991}
  (\bibinfo {year} {2013})}\BibitemShut {NoStop}%
\bibitem [{\citenamefont {Jiang}\ \emph {et~al.}(2013)\citenamefont {Jiang},
  \citenamefont {Li}, \citenamefont {Zhang}, \citenamefont {Sun}, \citenamefont
  {Chen}, \citenamefont {Ye}, \citenamefont {Xu}, \citenamefont {Ge},
  \citenamefont {Tan}, \citenamefont {Niu}, \citenamefont {Xia}, \citenamefont
  {Xie}, \citenamefont {Li}, \citenamefont {Chen}, \citenamefont {Wen},\ and\
  \citenamefont {Feng}}]{Jiang2013}%
  \BibitemOpen
  \bibfield  {author} {\bibinfo {author} {\bibfnamefont {J.}~\bibnamefont
  {Jiang}}, \bibinfo {author} {\bibfnamefont {S.}~\bibnamefont {Li}}, \bibinfo
  {author} {\bibfnamefont {T.}~\bibnamefont {Zhang}}, \bibinfo {author}
  {\bibfnamefont {Z.}~\bibnamefont {Sun}}, \bibinfo {author} {\bibfnamefont
  {F.}~\bibnamefont {Chen}}, \bibinfo {author} {\bibfnamefont {Z.R.}\
  \bibnamefont {Ye}}, \bibinfo {author} {\bibfnamefont {M.}~\bibnamefont {Xu}},
  \bibinfo {author} {\bibfnamefont {Q.Q.}\ \bibnamefont {Ge}}, \bibinfo
  {author} {\bibfnamefont {S.Y.}\ \bibnamefont {Tan}}, \bibinfo {author}
  {\bibfnamefont {X.H.}\ \bibnamefont {Niu}}, \bibinfo {author} {\bibfnamefont
  {M.}~\bibnamefont {Xia}}, \bibinfo {author} {\bibfnamefont {B.P.}\
  \bibnamefont {Xie}}, \bibinfo {author} {\bibfnamefont {Y.F.}\ \bibnamefont
  {Li}}, \bibinfo {author} {\bibfnamefont {X.H.}\ \bibnamefont {Chen}},
  \bibinfo {author} {\bibfnamefont {H.H.}\ \bibnamefont {Wen}}, \ and\ \bibinfo
  {author} {\bibfnamefont {D.L.}\ \bibnamefont {Feng}},\ }\bibfield  {title}
  {\enquote {\bibinfo {title} {Observation of possible topological in-gap
  surface states in the {K}ondo insulator {SmB}$_6$ by photoemission},}\ }\href
  {\doibase 10.1038/ncomms4010} {\bibfield  {journal} {\bibinfo  {journal}
  {Nature Comm.}\ }\textbf {\bibinfo {volume} {4}},\ \bibinfo {pages} {3010}
  (\bibinfo {year} {2013})}\BibitemShut {NoStop}%
\bibitem [{\citenamefont {Denlinger}\ \emph {et~al.}(2013)\citenamefont
  {Denlinger}, \citenamefont {Allen}, \citenamefont {Kang}, \citenamefont
  {Sund}, \citenamefont {Kim}, \citenamefont {Shim}, \citenamefont {Min},
  \citenamefont {Kim},\ and\ \citenamefont {Fisk}}]{Denlinger2013}%
  \BibitemOpen
  \bibfield  {author} {\bibinfo {author} {\bibfnamefont {J.~D.}\ \bibnamefont
  {Denlinger}}, \bibinfo {author} {\bibfnamefont {J.~W.}\ \bibnamefont
  {Allen}}, \bibinfo {author} {\bibfnamefont {J.-S.}\ \bibnamefont {Kang}},
  \bibinfo {author} {\bibfnamefont {K.}~\bibnamefont {Sund}}, \bibinfo {author}
  {\bibfnamefont {J.-W.}\ \bibnamefont {Kim}}, \bibinfo {author} {\bibfnamefont
  {J.~H.}\ \bibnamefont {Shim}}, \bibinfo {author} {\bibfnamefont {B.~I.}\
  \bibnamefont {Min}}, \bibinfo {author} {\bibfnamefont {D.-J.}\ \bibnamefont
  {Kim}}, \ and\ \bibinfo {author} {\bibfnamefont {Z.}~\bibnamefont {Fisk}},\
  }\bibfield  {title} {\enquote {\bibinfo {title} {Temperature dependence of
  linked gap and surface state evolution in the mixed valent topological
  insulator {SmB}$_6$},}\ }\href@noop {} {\bibfield  {journal} {\bibinfo
  {journal} {arXiv:1312.6637}\ } (\bibinfo {year} {2013})}\BibitemShut
  {NoStop}%
\bibitem [{\citenamefont {Denlinger}\ \emph {et~al.}(2014)\citenamefont
  {Denlinger}, \citenamefont {Allen}, \citenamefont {Kang}, \citenamefont
  {Sun}, \citenamefont {Min}, \citenamefont {Kim},\ and\ \citenamefont
  {Fisk}}]{Denlinger2014}%
  \BibitemOpen
  \bibfield  {author} {\bibinfo {author} {\bibfnamefont {J.~D.}\ \bibnamefont
  {Denlinger}}, \bibinfo {author} {\bibfnamefont {J.~W.}\ \bibnamefont
  {Allen}}, \bibinfo {author} {\bibfnamefont {J.-S.}\ \bibnamefont {Kang}},
  \bibinfo {author} {\bibfnamefont {K.}~\bibnamefont {Sun}}, \bibinfo {author}
  {\bibfnamefont {B.-II}\ \bibnamefont {Min}}, \bibinfo {author} {\bibfnamefont
  {D.~J-.}\ \bibnamefont {Kim}}, \ and\ \bibinfo {author} {\bibfnamefont
  {Z.}~\bibnamefont {Fisk}},\ }\bibfield  {title} {\enquote {\bibinfo {title}
  {{SmB}$_6$ {P}hotoemission: {P}ast and {P}resent},}\ }\href {\doibase
  10.7566/JPSCP.3.017038} {\bibfield  {journal} {\bibinfo  {journal} {JPS Conf.
  Proc.}\ }\textbf {\bibinfo {volume} {3}},\ \bibinfo {pages} {017038}
  (\bibinfo {year} {2014})}\BibitemShut {NoStop}%
\bibitem [{\citenamefont {Hatnean}\ \emph {et~al.}(2013)\citenamefont
  {Hatnean}, \citenamefont {Lees}, \citenamefont {Paul},\ and\ \citenamefont
  {Balakrishnan}}]{Hatnean2013}%
  \BibitemOpen
  \bibfield  {author} {\bibinfo {author} {\bibfnamefont {M.~C.}\ \bibnamefont
  {Hatnean}}, \bibinfo {author} {\bibfnamefont {M.~R.}\ \bibnamefont {Lees}},
  \bibinfo {author} {\bibfnamefont {D.~M.~K.}\ \bibnamefont {Paul}}, \ and\
  \bibinfo {author} {\bibfnamefont {G.}~\bibnamefont {Balakrishnan}},\
  }\bibfield  {title} {\enquote {\bibinfo {title} {Large, high quality
  single-crystals of the new topological {K}ondo insulator, {SmB}$_6$},}\
  }\href {\doibase doi:10.1038/srep03071} {\bibfield  {journal} {\bibinfo
  {journal} {Sci. Rep.}\ ,\ \bibinfo {pages} {3403}} (\bibinfo {year}
  {2013})}\BibitemShut {NoStop}%
\bibitem [{\citenamefont {Zhang}\ \emph {et~al.}(2013)\citenamefont {Zhang},
  \citenamefont {Butch}, \citenamefont {Syers}, \citenamefont {Ziemak},
  \citenamefont {Greene},\ and\ \citenamefont {Paglione}}]{Zhang2013}%
  \BibitemOpen
  \bibfield  {author} {\bibinfo {author} {\bibfnamefont {X.}~\bibnamefont
  {Zhang}}, \bibinfo {author} {\bibfnamefont {N.~P.}\ \bibnamefont {Butch}},
  \bibinfo {author} {\bibfnamefont {P.}~\bibnamefont {Syers}}, \bibinfo
  {author} {\bibfnamefont {S.}~\bibnamefont {Ziemak}}, \bibinfo {author}
  {\bibfnamefont {R.~L.}\ \bibnamefont {Greene}}, \ and\ \bibinfo {author}
  {\bibfnamefont {J.}~\bibnamefont {Paglione}},\ }\bibfield  {title} {\enquote
  {\bibinfo {title} {Hybridization, inter-ion correlation, and surface states
  in the {K}ondo insulator {SmB}$_6$},}\ }\href {\doibase
  10.1103/PhysRevX.3.011011} {\bibfield  {journal} {\bibinfo  {journal} {Phys.
  Rev. X}\ }\textbf {\bibinfo {volume} {3}},\ \bibinfo {pages} {011011}
  (\bibinfo {year} {2013})}\BibitemShut {NoStop}%
\bibitem [{\citenamefont {Kim}\ \emph {et~al.}(2013)\citenamefont {Kim},
  \citenamefont {Thomas}, \citenamefont {Grant}, \citenamefont {Botimer},
  \citenamefont {Fisk},\ and\ \citenamefont {Xia}}]{Kim2013}%
  \BibitemOpen
  \bibfield  {author} {\bibinfo {author} {\bibfnamefont {D.~J.}\ \bibnamefont
  {Kim}}, \bibinfo {author} {\bibfnamefont {S.}~\bibnamefont {Thomas}},
  \bibinfo {author} {\bibfnamefont {T.}~\bibnamefont {Grant}}, \bibinfo
  {author} {\bibfnamefont {J.}~\bibnamefont {Botimer}}, \bibinfo {author}
  {\bibfnamefont {Z.}~\bibnamefont {Fisk}}, \ and\ \bibinfo {author}
  {\bibfnamefont {J.}~\bibnamefont {Xia}},\ }\bibfield  {title} {\enquote
  {\bibinfo {title} {Surface hall effect and nonlocal transport in {SmB}$_6$:
  Evidence for surface conduction},}\ }\href {\doibase 10.1038/srep03150}
  {\bibfield  {journal} {\bibinfo  {journal} {Sci. Rep.}\ }\textbf {\bibinfo
  {volume} {3}},\ \bibinfo {pages} {3150} (\bibinfo {year} {2013})}\BibitemShut
  {NoStop}%
\bibitem [{\citenamefont {Wolgast}\ \emph {et~al.}(2013)\citenamefont
  {Wolgast}, \citenamefont {Kurdak}, \citenamefont {Sun}, \citenamefont
  {Allen}, \citenamefont {Kim},\ and\ \citenamefont {Fisk}}]{Wolgast2013}%
  \BibitemOpen
  \bibfield  {author} {\bibinfo {author} {\bibfnamefont {S.}~\bibnamefont
  {Wolgast}}, \bibinfo {author} {\bibfnamefont {C.}~\bibnamefont {Kurdak}},
  \bibinfo {author} {\bibfnamefont {K.}~\bibnamefont {Sun}}, \bibinfo {author}
  {\bibfnamefont {J.~W.}\ \bibnamefont {Allen}}, \bibinfo {author}
  {\bibfnamefont {D.~J.}\ \bibnamefont {Kim}}, \ and\ \bibinfo {author}
  {\bibfnamefont {Z.}~\bibnamefont {Fisk}},\ }\bibfield  {title} {\enquote
  {\bibinfo {title} {Low-temperature surface conduction in the {K}ondo
  insulator {SmB}$_6$},}\ }\href {\doibase 10.1103/PhysRevB.88.180405}
  {\bibfield  {journal} {\bibinfo  {journal} {Phys. Rev. B}\ }\textbf {\bibinfo
  {volume} {88}},\ \bibinfo {pages} {180405} (\bibinfo {year}
  {2013})}\BibitemShut {NoStop}%
\bibitem [{\citenamefont {Kim}\ \emph {et~al.}(2014{\natexlab{a}})\citenamefont
  {Kim}, \citenamefont {Xia},\ and\ \citenamefont {Fisk}}]{Kim2014}%
  \BibitemOpen
  \bibfield  {author} {\bibinfo {author} {\bibfnamefont {D.~J.}\ \bibnamefont
  {Kim}}, \bibinfo {author} {\bibfnamefont {J.}~\bibnamefont {Xia}}, \ and\
  \bibinfo {author} {\bibfnamefont {Z.}~\bibnamefont {Fisk}},\ }\bibfield
  {title} {\enquote {\bibinfo {title} {Topological surface state in the {Kondo}
  insulator samarium hexaboride},}\ }\href {\doibase
  10.1103/PhysRevLett.22.295} {\bibfield  {journal} {\bibinfo  {journal} {Nat.
  Mater.}\ }\textbf {\bibinfo {volume} {13}},\ \bibinfo {pages} {466--470}
  (\bibinfo {year} {2014}{\natexlab{a}})}\BibitemShut {NoStop}%
\bibitem [{\citenamefont {Wolgast}\ \emph {et~al.}(2015)\citenamefont
  {Wolgast}, \citenamefont {Eo}, \citenamefont {\"Ozt\"urk}, \citenamefont
  {Li}, \citenamefont {Xiang}, \citenamefont {Tinsman}, \citenamefont {Asaba},
  \citenamefont {Lawson}, \citenamefont {Yu}, \citenamefont {Allen},
  \citenamefont {Sun}, \citenamefont {Li}, \citenamefont {Kurdak},
  \citenamefont {Kim},\ and\ \citenamefont {Fisk}}]{Wolgast2015}%
  \BibitemOpen
  \bibfield  {author} {\bibinfo {author} {\bibfnamefont {S.}~\bibnamefont
  {Wolgast}}, \bibinfo {author} {\bibfnamefont {Y.~S.}\ \bibnamefont {Eo}},
  \bibinfo {author} {\bibfnamefont {T.}~\bibnamefont {\"Ozt\"urk}}, \bibinfo
  {author} {\bibfnamefont {G.}~\bibnamefont {Li}}, \bibinfo {author}
  {\bibfnamefont {Z.}~\bibnamefont {Xiang}}, \bibinfo {author} {\bibfnamefont
  {C.}~\bibnamefont {Tinsman}}, \bibinfo {author} {\bibfnamefont
  {T.}~\bibnamefont {Asaba}}, \bibinfo {author} {\bibfnamefont
  {B.}~\bibnamefont {Lawson}}, \bibinfo {author} {\bibfnamefont
  {F.}~\bibnamefont {Yu}}, \bibinfo {author} {\bibfnamefont {J.~W.}\
  \bibnamefont {Allen}}, \bibinfo {author} {\bibfnamefont {K.}~\bibnamefont
  {Sun}}, \bibinfo {author} {\bibfnamefont {L.}~\bibnamefont {Li}}, \bibinfo
  {author} {\bibfnamefont {C.}~\bibnamefont {Kurdak}}, \bibinfo {author}
  {\bibfnamefont {D.-J.}\ \bibnamefont {Kim}}, \ and\ \bibinfo {author}
  {\bibfnamefont {Z.}~\bibnamefont {Fisk}},\ }\bibfield  {title} {\enquote
  {\bibinfo {title} {Magnetotransport measurements of the surface states of
  samarium hexaboride using corbino structures},}\ }\href {\doibase
  10.1103/PhysRevB.92.115110} {\bibfield  {journal} {\bibinfo  {journal} {Phys.
  Rev. B}\ }\textbf {\bibinfo {volume} {92}},\ \bibinfo {pages} {115110}
  (\bibinfo {year} {2015})}\BibitemShut {NoStop}%
\bibitem [{\citenamefont {Thomas}\ \emph {et~al.}(2016)\citenamefont {Thomas},
  \citenamefont {Kim}, \citenamefont {Chung}, \citenamefont {Grant},
  \citenamefont {Fisk},\ and\ \citenamefont {Xia}}]{Thomas2016}%
  \BibitemOpen
  \bibfield  {author} {\bibinfo {author} {\bibfnamefont {S.}~\bibnamefont
  {Thomas}}, \bibinfo {author} {\bibfnamefont {D.~J.}\ \bibnamefont {Kim}},
  \bibinfo {author} {\bibfnamefont {S.~B.}\ \bibnamefont {Chung}}, \bibinfo
  {author} {\bibfnamefont {T.}~\bibnamefont {Grant}}, \bibinfo {author}
  {\bibfnamefont {Z.}~\bibnamefont {Fisk}}, \ and\ \bibinfo {author}
  {\bibfnamefont {Jing}\ \bibnamefont {Xia}},\ }\bibfield  {title} {\enquote
  {\bibinfo {title} {Weak antilocalization and linear magnetoresistance in the
  surface state of {SmB}$_6$},}\ }\href {\doibase 10.1103/PhysRevB.94.205114}
  {\bibfield  {journal} {\bibinfo  {journal} {Phys. Rev. B}\ }\textbf {\bibinfo
  {volume} {94}},\ \bibinfo {pages} {205114} (\bibinfo {year}
  {2016})}\BibitemShut {NoStop}%
\bibitem [{\citenamefont {Nakajima}\ \emph {et~al.}(2016)\citenamefont
  {Nakajima}, \citenamefont {Syers}, \citenamefont {Wang}, \citenamefont
  {Wang},\ and\ \citenamefont {Paglione}}]{Nakajima2016}%
  \BibitemOpen
  \bibfield  {author} {\bibinfo {author} {\bibfnamefont {Y.}~\bibnamefont
  {Nakajima}}, \bibinfo {author} {\bibfnamefont {P.}~\bibnamefont {Syers}},
  \bibinfo {author} {\bibfnamefont {X.}~\bibnamefont {Wang}}, \bibinfo {author}
  {\bibfnamefont {R.}~\bibnamefont {Wang}}, \ and\ \bibinfo {author}
  {\bibfnamefont {J.}~\bibnamefont {Paglione}},\ }\bibfield  {title} {\enquote
  {\bibinfo {title} {One-dimensional edge state transport in a topological
  {K}ondo insulator},}\ }\href {\doibase 10.1038/nphys3555} {\bibfield
  {journal} {\bibinfo  {journal} {Nature Phys.}\ }\textbf {\bibinfo {volume}
  {12}},\ \bibinfo {pages} {213--217} (\bibinfo {year} {2016})}\BibitemShut
  {NoStop}%
\bibitem [{\citenamefont {Dzero}\ \emph {et~al.}(2010)\citenamefont {Dzero},
  \citenamefont {Sun}, \citenamefont {Galitski},\ and\ \citenamefont
  {Coleman}}]{Dzero2010}%
  \BibitemOpen
  \bibfield  {author} {\bibinfo {author} {\bibfnamefont {M.}~\bibnamefont
  {Dzero}}, \bibinfo {author} {\bibfnamefont {K.}~\bibnamefont {Sun}}, \bibinfo
  {author} {\bibfnamefont {V.}~\bibnamefont {Galitski}}, \ and\ \bibinfo
  {author} {\bibfnamefont {P.}~\bibnamefont {Coleman}},\ }\bibfield  {title}
  {\enquote {\bibinfo {title} {Topological {K}ondo insulators},}\ }\href
  {\doibase 10.1103/PhysRevLett.104.106408} {\bibfield  {journal} {\bibinfo
  {journal} {Phys. Rev. Lett.}\ }\textbf {\bibinfo {volume} {104}},\ \bibinfo
  {pages} {106408} (\bibinfo {year} {2010})}\BibitemShut {NoStop}%
\bibitem [{\citenamefont {Takimoto}(2011)}]{Takimoto2011}%
  \BibitemOpen
  \bibfield  {author} {\bibinfo {author} {\bibfnamefont {T.}~\bibnamefont
  {Takimoto}},\ }\bibfield  {title} {\enquote {\bibinfo {title} {Sm{B}$_6$: {A}
  promising candidate for a topological insulator},}\ }\href {\doibase
  10.1143/JPSJ.80.123710} {\bibfield  {journal} {\bibinfo  {journal} {J. Phys.
  Soc. Jpn.}\ }\textbf {\bibinfo {volume} {80}},\ \bibinfo {pages} {123710}
  (\bibinfo {year} {2011})}\BibitemShut {NoStop}%
\bibitem [{\citenamefont {Lu}\ \emph {et~al.}(2013)\citenamefont {Lu},
  \citenamefont {Zhao}, \citenamefont {Weng}, \citenamefont {Fang},\ and\
  \citenamefont {Dai}}]{Lu2013}%
  \BibitemOpen
  \bibfield  {author} {\bibinfo {author} {\bibfnamefont {F.}~\bibnamefont
  {Lu}}, \bibinfo {author} {\bibfnamefont {J.~Z.}\ \bibnamefont {Zhao}},
  \bibinfo {author} {\bibfnamefont {H.}~\bibnamefont {Weng}}, \bibinfo {author}
  {\bibfnamefont {Z.}~\bibnamefont {Fang}}, \ and\ \bibinfo {author}
  {\bibfnamefont {X.}~\bibnamefont {Dai}},\ }\bibfield  {title} {\enquote
  {\bibinfo {title} {Correlated topological insulators with mixed valence},}\
  }\href {\doibase 10.1103/PhysRevLett.110.096401} {\bibfield  {journal}
  {\bibinfo  {journal} {Phys. Rev. Lett.}\ }\textbf {\bibinfo {volume} {110}},\
  \bibinfo {pages} {096401} (\bibinfo {year} {2013})}\BibitemShut {NoStop}%
\bibitem [{\citenamefont {Alexandrov}\ \emph {et~al.}(2013)\citenamefont
  {Alexandrov}, \citenamefont {Dzero},\ and\ \citenamefont
  {Coleman}}]{Alexandrov2013}%
  \BibitemOpen
  \bibfield  {author} {\bibinfo {author} {\bibfnamefont {V.}~\bibnamefont
  {Alexandrov}}, \bibinfo {author} {\bibfnamefont {M.}~\bibnamefont {Dzero}}, \
  and\ \bibinfo {author} {\bibfnamefont {P.}~\bibnamefont {Coleman}},\
  }\bibfield  {title} {\enquote {\bibinfo {title} {Cubic topological {K}ondo
  insulators},}\ }\href {\doibase 10.1103/PhysRevLett.111.226403} {\bibfield
  {journal} {\bibinfo  {journal} {Phys. Rev. Lett.}\ }\textbf {\bibinfo
  {volume} {111}},\ \bibinfo {pages} {226403} (\bibinfo {year}
  {2013})}\BibitemShut {NoStop}%
\bibitem [{\citenamefont {Frantzeskakis}\ \emph {et~al.}(2013)\citenamefont
  {Frantzeskakis}, \citenamefont {de~Jong}, \citenamefont {Zwartsenberg},
  \citenamefont {Huang}, \citenamefont {Pan}, \citenamefont {Zhang},
  \citenamefont {Zhang}, \citenamefont {Zhang}, \citenamefont {Bao},
  \citenamefont {Tegus}, \citenamefont {Varykhalov}, \citenamefont
  {de~Visser},\ and\ \citenamefont {Golden}}]{Frantzeskakis2013}%
  \BibitemOpen
  \bibfield  {author} {\bibinfo {author} {\bibfnamefont {E.}~\bibnamefont
  {Frantzeskakis}}, \bibinfo {author} {\bibfnamefont {N.}~\bibnamefont
  {de~Jong}}, \bibinfo {author} {\bibfnamefont {B.}~\bibnamefont
  {Zwartsenberg}}, \bibinfo {author} {\bibfnamefont {Y.~K.}\ \bibnamefont
  {Huang}}, \bibinfo {author} {\bibfnamefont {Y.}~\bibnamefont {Pan}}, \bibinfo
  {author} {\bibfnamefont {X.}~\bibnamefont {Zhang}}, \bibinfo {author}
  {\bibfnamefont {J.~X.}\ \bibnamefont {Zhang}}, \bibinfo {author}
  {\bibfnamefont {F.~X.}\ \bibnamefont {Zhang}}, \bibinfo {author}
  {\bibfnamefont {L.~H.}\ \bibnamefont {Bao}}, \bibinfo {author} {\bibfnamefont
  {O.}~\bibnamefont {Tegus}}, \bibinfo {author} {\bibfnamefont
  {A.}~\bibnamefont {Varykhalov}}, \bibinfo {author} {\bibfnamefont
  {A.}~\bibnamefont {de~Visser}}, \ and\ \bibinfo {author} {\bibfnamefont
  {M.~S.}\ \bibnamefont {Golden}},\ }\bibfield  {title} {\enquote {\bibinfo
  {title} {Kondo hybridization and the origin of metallic states at the (001)
  surface of {SmB}$_6$},}\ }\href {\doibase 10.1103/PhysRevX.3.041024}
  {\bibfield  {journal} {\bibinfo  {journal} {Phys. Rev. X}\ }\textbf {\bibinfo
  {volume} {3(4)}},\ \bibinfo {pages} {041024} (\bibinfo {year}
  {2013})}\BibitemShut {NoStop}%
\bibitem [{\citenamefont {Xu}\ \emph {et~al.}(2014{\natexlab{a}})\citenamefont
  {Xu}, \citenamefont {Matt}, \citenamefont {Pomjakushina}, \citenamefont
  {Shi}, \citenamefont {Dhaka}, \citenamefont {Plumb}, \citenamefont
  {Radovi\ifmmode~\acute{c}\else \'{c}\fi{}}, \citenamefont {Biswas},
  \citenamefont {Evtushinsky}, \citenamefont {Zabolotnyy}, \citenamefont {Dil},
  \citenamefont {Conder}, \citenamefont {Mesot}, \citenamefont {Ding},\ and\
  \citenamefont {Shi}}]{Xu2014}%
  \BibitemOpen
  \bibfield  {author} {\bibinfo {author} {\bibfnamefont {N.}~\bibnamefont
  {Xu}}, \bibinfo {author} {\bibfnamefont {C.~E.}\ \bibnamefont {Matt}},
  \bibinfo {author} {\bibfnamefont {E.}~\bibnamefont {Pomjakushina}}, \bibinfo
  {author} {\bibfnamefont {X.}~\bibnamefont {Shi}}, \bibinfo {author}
  {\bibfnamefont {R.~S.}\ \bibnamefont {Dhaka}}, \bibinfo {author}
  {\bibfnamefont {N.~C.}\ \bibnamefont {Plumb}}, \bibinfo {author}
  {\bibfnamefont {M.}~\bibnamefont {Radovi\ifmmode~\acute{c}\else \'{c}\fi{}}},
  \bibinfo {author} {\bibfnamefont {P.~K.}\ \bibnamefont {Biswas}}, \bibinfo
  {author} {\bibfnamefont {D.}~\bibnamefont {Evtushinsky}}, \bibinfo {author}
  {\bibfnamefont {V.}~\bibnamefont {Zabolotnyy}}, \bibinfo {author}
  {\bibfnamefont {J.~H.}\ \bibnamefont {Dil}}, \bibinfo {author} {\bibfnamefont
  {K.}~\bibnamefont {Conder}}, \bibinfo {author} {\bibfnamefont
  {J.}~\bibnamefont {Mesot}}, \bibinfo {author} {\bibfnamefont
  {H.}~\bibnamefont {Ding}}, \ and\ \bibinfo {author} {\bibfnamefont
  {M.}~\bibnamefont {Shi}},\ }\bibfield  {title} {\enquote {\bibinfo {title}
  {Exotic {K}ondo crossover in a wide temperature region in the topological
  {K}ondo insulator {SmB}$_6$ revealed by high-resolution {ARPES}},}\ }\href
  {\doibase 10.1103/PhysRevB.90.085148} {\bibfield  {journal} {\bibinfo
  {journal} {Phys. Rev. B}\ }\textbf {\bibinfo {volume} {90}},\ \bibinfo
  {pages} {085148} (\bibinfo {year} {2014}{\natexlab{a}})}\BibitemShut
  {NoStop}%
\bibitem [{\citenamefont {Xu}\ \emph {et~al.}(2014{\natexlab{b}})\citenamefont
  {Xu}, \citenamefont {Biswas}, \citenamefont {Dil}, \citenamefont {Dhaka},
  \citenamefont {Landolt}, \citenamefont {Muff}, \citenamefont {Matt},
  \citenamefont {Shi}, \citenamefont {Plumb}, \citenamefont {Radovic},
  \citenamefont {Pomjakushina}, \citenamefont {Conder}, \citenamefont {Amato},
  \citenamefont {Borisenko}, \citenamefont {Yu}, \citenamefont {Weng},
  \citenamefont {Fang}, \citenamefont {Dai}, \citenamefont {Mesot},
  \citenamefont {Ding},\ and\ \citenamefont {Shi}}]{Xu2014natcomm}%
  \BibitemOpen
  \bibfield  {author} {\bibinfo {author} {\bibfnamefont {N.}~\bibnamefont
  {Xu}}, \bibinfo {author} {\bibfnamefont {P.~K.}\ \bibnamefont {Biswas}},
  \bibinfo {author} {\bibfnamefont {J.~H.}\ \bibnamefont {Dil}}, \bibinfo
  {author} {\bibfnamefont {R.~S.}\ \bibnamefont {Dhaka}}, \bibinfo {author}
  {\bibfnamefont {G.}~\bibnamefont {Landolt}}, \bibinfo {author} {\bibfnamefont
  {S.}~\bibnamefont {Muff}}, \bibinfo {author} {\bibfnamefont {C.~E.}\
  \bibnamefont {Matt}}, \bibinfo {author} {\bibfnamefont {X.}~\bibnamefont
  {Shi}}, \bibinfo {author} {\bibfnamefont {N.~C.}\ \bibnamefont {Plumb}},
  \bibinfo {author} {\bibfnamefont {M.}~\bibnamefont {Radovic}}, \bibinfo
  {author} {\bibfnamefont {E.}~\bibnamefont {Pomjakushina}}, \bibinfo {author}
  {\bibfnamefont {K.}~\bibnamefont {Conder}}, \bibinfo {author} {\bibfnamefont
  {A.}~\bibnamefont {Amato}}, \bibinfo {author} {\bibfnamefont {S.~V.}\
  \bibnamefont {Borisenko}}, \bibinfo {author} {\bibfnamefont {R.}~\bibnamefont
  {Yu}}, \bibinfo {author} {\bibfnamefont {H.-M.}\ \bibnamefont {Weng}},
  \bibinfo {author} {\bibfnamefont {Z.}~\bibnamefont {Fang}}, \bibinfo {author}
  {\bibfnamefont {X.}~\bibnamefont {Dai}}, \bibinfo {author} {\bibfnamefont
  {J.}~\bibnamefont {Mesot}}, \bibinfo {author} {\bibfnamefont
  {H.}~\bibnamefont {Ding}}, \ and\ \bibinfo {author} {\bibfnamefont
  {M.}~\bibnamefont {Shi}},\ }\bibfield  {title} {\enquote {\bibinfo {title}
  {Direct observation of the spin texture in {SmB$_6$} as evidence of the
  topological {Kondo} insulator},}\ }\href {\doibase 10.1038/ncomms5566}
  {\bibfield  {journal} {\bibinfo  {journal} {Nat. Commun.}\ }\textbf {\bibinfo
  {volume} {5}},\ \bibinfo {pages} {4566} (\bibinfo {year}
  {2014}{\natexlab{b}})}\BibitemShut {NoStop}%
\bibitem [{\citenamefont {Hlawenka}\ \emph {et~al.}(2018)\citenamefont
  {Hlawenka}, \citenamefont {Siemensmeyer}, \citenamefont {Weschke},
  \citenamefont {Varykhalov}, \citenamefont {Sánchez-Barriga}, \citenamefont
  {Shitsevalova}, \citenamefont {Dukhnenko}, \citenamefont {Filipov},
  \citenamefont {Gabáni}, \citenamefont {Flachbart}, \citenamefont {Rader},\
  and\ \citenamefont {Rienks}}]{Hlawenka2015}%
  \BibitemOpen
  \bibfield  {author} {\bibinfo {author} {\bibfnamefont {P.}~\bibnamefont
  {Hlawenka}}, \bibinfo {author} {\bibfnamefont {K.}~\bibnamefont
  {Siemensmeyer}}, \bibinfo {author} {\bibfnamefont {E.}~\bibnamefont
  {Weschke}}, \bibinfo {author} {\bibfnamefont {A.}~\bibnamefont {Varykhalov}},
  \bibinfo {author} {\bibfnamefont {J.}~\bibnamefont {Sánchez-Barriga}},
  \bibinfo {author} {\bibfnamefont {N.~Y.}\ \bibnamefont {Shitsevalova}},
  \bibinfo {author} {\bibfnamefont {A.~V.}\ \bibnamefont {Dukhnenko}}, \bibinfo
  {author} {\bibfnamefont {V.~B.}\ \bibnamefont {Filipov}}, \bibinfo {author}
  {\bibfnamefont {S.}~\bibnamefont {Gabáni}}, \bibinfo {author} {\bibfnamefont
  {S.}~\bibnamefont {Flachbart}}, \bibinfo {author} {\bibfnamefont
  {O.}~\bibnamefont {Rader}}, \ and\ \bibinfo {author} {\bibfnamefont
  {E.~D.~L.}\ \bibnamefont {Rienks}},\ }\bibfield  {title} {\enquote {\bibinfo
  {title} {Samarium hexaboride: A trivial surface conductor},}\ }\href
  {\doibase 10.1038/s41467-018-02908-7} {\bibfield  {journal} {\bibinfo
  {journal} {Nat. Comm.}\ }\textbf {\bibinfo {volume} {9}},\ \bibinfo {pages}
  {517} (\bibinfo {year} {2018})}\BibitemShut {NoStop}%
\bibitem [{\citenamefont {Ohtsubo}\ \emph {et~al.}(2019)\citenamefont
  {Ohtsubo}, \citenamefont {Yamashita}, \citenamefont {Hagiwara}, \citenamefont
  {Ideta}, \citenamefont {Tanaka}, \citenamefont {Yukawa}, \citenamefont
  {Horiba}, \citenamefont {Kumigashira}, \citenamefont {Miyamoto},
  \citenamefont {Okuda}, \citenamefont {Hirano}, \citenamefont {Iga},\ and\
  \citenamefont {Kimura}}]{Ohtsubo2019}%
  \BibitemOpen
  \bibfield  {author} {\bibinfo {author} {\bibfnamefont {Y.}~\bibnamefont
  {Ohtsubo}}, \bibinfo {author} {\bibfnamefont {Y.}~\bibnamefont {Yamashita}},
  \bibinfo {author} {\bibfnamefont {K.}~\bibnamefont {Hagiwara}}, \bibinfo
  {author} {\bibfnamefont {S.}~\bibnamefont {Ideta}}, \bibinfo {author}
  {\bibfnamefont {K.}~\bibnamefont {Tanaka}}, \bibinfo {author} {\bibfnamefont
  {R.}~\bibnamefont {Yukawa}}, \bibinfo {author} {\bibfnamefont
  {K.}~\bibnamefont {Horiba}}, \bibinfo {author} {\bibfnamefont
  {H.}~\bibnamefont {Kumigashira}}, \bibinfo {author} {\bibfnamefont
  {K.}~\bibnamefont {Miyamoto}}, \bibinfo {author} {\bibfnamefont
  {T.}~\bibnamefont {Okuda}}, \bibinfo {author} {\bibfnamefont
  {W.}~\bibnamefont {Hirano}}, \bibinfo {author} {\bibfnamefont
  {F.}~\bibnamefont {Iga}}, \ and\ \bibinfo {author} {\bibfnamefont
  {S.}~\bibnamefont {Kimura}},\ }\bibfield  {title} {\enquote {\bibinfo {title}
  {Non-trivial surface states of samarium hexaboride at the (111) surface},}\
  }\href {\doibase 10.1038/s41467-019-10353-3} {\bibfield  {journal} {\bibinfo
  {journal} {Nat. Comm.}\ }\textbf {\bibinfo {volume} {10}},\ \bibinfo {pages}
  {2298} (\bibinfo {year} {2019})}\BibitemShut {NoStop}%
\bibitem [{\citenamefont {Ruan}\ \emph {et~al.}(2014)\citenamefont {Ruan},
  \citenamefont {Ye}, \citenamefont {Guo}, \citenamefont {Chen}, \citenamefont
  {Chen}, \citenamefont {Zhang},\ and\ \citenamefont {Wang}}]{Ruan2014}%
  \BibitemOpen
  \bibfield  {author} {\bibinfo {author} {\bibfnamefont {W.}~\bibnamefont
  {Ruan}}, \bibinfo {author} {\bibfnamefont {C.}~\bibnamefont {Ye}}, \bibinfo
  {author} {\bibfnamefont {M.}~\bibnamefont {Guo}}, \bibinfo {author}
  {\bibfnamefont {F.}~\bibnamefont {Chen}}, \bibinfo {author} {\bibfnamefont
  {X.}~\bibnamefont {Chen}}, \bibinfo {author} {\bibfnamefont {G..-M.}\
  \bibnamefont {Zhang}}, \ and\ \bibinfo {author} {\bibfnamefont
  {Y.}~\bibnamefont {Wang}},\ }\bibfield  {title} {\enquote {\bibinfo {title}
  {Emergence of a coherent in-gap state in the {SmB}$_6$ {K}ondo insulator
  revealed by scanning tunneling spectroscopy},}\ }\href {\doibase
  10.1103/PhysRevLett.112.136401} {\bibfield  {journal} {\bibinfo  {journal}
  {Phys. Rev. Lett.}\ }\textbf {\bibinfo {volume} {112}},\ \bibinfo {pages}
  {136401} (\bibinfo {year} {2014})}\BibitemShut {NoStop}%
\bibitem [{\citenamefont {R\"ossler}\ \emph {et~al.}(2014)\citenamefont
  {R\"ossler}, \citenamefont {Jang}, \citenamefont {Kim}, \citenamefont {H.},
  \citenamefont {Fisk},\ and\ \citenamefont {Steglich}}]{Roessler2014}%
  \BibitemOpen
  \bibfield  {author} {\bibinfo {author} {\bibfnamefont {S.}~\bibnamefont
  {R\"ossler}}, \bibinfo {author} {\bibfnamefont {T.~H.}\ \bibnamefont {Jang}},
  \bibinfo {author} {\bibfnamefont {D.~J.}\ \bibnamefont {Kim}}, \bibinfo
  {author} {\bibfnamefont {Tjeng.~L.}\ \bibnamefont {H.}}, \bibinfo {author}
  {\bibfnamefont {Z.}~\bibnamefont {Fisk}}, \ and\ \bibinfo {author}
  {\bibfnamefont {S}~\bibnamefont {Steglich}, \bibfnamefont {F.~Wirth}},\
  }\bibfield  {title} {\enquote {\bibinfo {title} {Hybridization gap and fano
  resonance in {SmB}$_6$},}\ }\href {\doibase 10.1073/pnas.1402643111}
  {\bibfield  {journal} {\bibinfo  {journal} {Proc. Nat. Acad. Science.
  U.S.A.}\ }\textbf {\bibinfo {volume} {111}},\ \bibinfo {pages} {4798}
  (\bibinfo {year} {2014})}\BibitemShut {NoStop}%
\bibitem [{\citenamefont {R\"ossler}\ \emph {et~al.}(2016)\citenamefont
  {R\"ossler}, \citenamefont {Jiao}, \citenamefont {Kim}, \citenamefont
  {Seiro}, \citenamefont {Rasim}, \citenamefont {Steglich}, \citenamefont
  {Tjeng}, \citenamefont {Fisk},\ and\ \citenamefont {Wirth}}]{Roessler2016}%
  \BibitemOpen
  \bibfield  {author} {\bibinfo {author} {\bibfnamefont {S.}~\bibnamefont
  {R\"ossler}}, \bibinfo {author} {\bibfnamefont {Lin}\ \bibnamefont {Jiao}},
  \bibinfo {author} {\bibfnamefont {D.~J.}\ \bibnamefont {Kim}}, \bibinfo
  {author} {\bibfnamefont {S.}~\bibnamefont {Seiro}}, \bibinfo {author}
  {\bibfnamefont {K.}~\bibnamefont {Rasim}}, \bibinfo {author} {\bibfnamefont
  {F.}~\bibnamefont {Steglich}}, \bibinfo {author} {\bibfnamefont {L.~H.}\
  \bibnamefont {Tjeng}}, \bibinfo {author} {\bibfnamefont {Z.}~\bibnamefont
  {Fisk}}, \ and\ \bibinfo {author} {\bibfnamefont {S.}~\bibnamefont {Wirth}},\
  }\bibfield  {title} {\enquote {\bibinfo {title} {Surface and electronic
  structure of {SmB}$_6$ through scanning tunneling microscopy},}\ }\href
  {\doibase 10.1080/14786435.2016.1171414} {\bibfield  {journal} {\bibinfo
  {journal} {Phil. Mag.}\ }\textbf {\bibinfo {volume} {96}},\ \bibinfo {pages}
  {3262--3273} (\bibinfo {year} {2016})}\BibitemShut {NoStop}%
\bibitem [{\citenamefont {Jiao}\ \emph {et~al.}(2016)\citenamefont {Jiao},
  \citenamefont {R\"ossler}, \citenamefont {Kim}, \citenamefont {Tjeng},
  \citenamefont {Fisk}, \citenamefont {Steglich},\ and\ \citenamefont
  {Wirth}}]{Jiao2016}%
  \BibitemOpen
  \bibfield  {author} {\bibinfo {author} {\bibfnamefont {L.}~\bibnamefont
  {Jiao}}, \bibinfo {author} {\bibfnamefont {S.}~\bibnamefont {R\"ossler}},
  \bibinfo {author} {\bibfnamefont {D.~J.}\ \bibnamefont {Kim}}, \bibinfo
  {author} {\bibfnamefont {L.~H.}\ \bibnamefont {Tjeng}}, \bibinfo {author}
  {\bibfnamefont {Z.}~\bibnamefont {Fisk}}, \bibinfo {author} {\bibfnamefont
  {F.}~\bibnamefont {Steglich}}, \ and\ \bibinfo {author} {\bibfnamefont
  {S.}~\bibnamefont {Wirth}},\ }\bibfield  {title} {\enquote {\bibinfo {title}
  {Hybridization gap and fano resonance in {SmB}$_6$},}\ }\href {\doibase
  10.1038/ncomms13762} {\bibfield  {journal} {\bibinfo  {journal} {Nat.
  Commun.}\ }\textbf {\bibinfo {volume} {7}},\ \bibinfo {pages} {13762}
  (\bibinfo {year} {2016})}\BibitemShut {NoStop}%
\bibitem [{\citenamefont {Li}\ \emph {et~al.}(2014)\citenamefont {Li},
  \citenamefont {Li}, \citenamefont {Blaha},\ and\ \citenamefont
  {Kioussis}}]{Li2014}%
  \BibitemOpen
  \bibfield  {author} {\bibinfo {author} {\bibfnamefont {Z.}~\bibnamefont
  {Li}}, \bibinfo {author} {\bibfnamefont {J.}~\bibnamefont {Li}}, \bibinfo
  {author} {\bibfnamefont {P.}~\bibnamefont {Blaha}}, \ and\ \bibinfo {author}
  {\bibfnamefont {N.}~\bibnamefont {Kioussis}},\ }\bibfield  {title} {\enquote
  {\bibinfo {title} {Predicted topological phase transition in the {SmS}
  {K}ondo insulator under pressure},}\ }\href {\doibase
  10.1103/PhysRevB.89.121117} {\bibfield  {journal} {\bibinfo  {journal} {Phys.
  Rev. B}\ }\textbf {\bibinfo {volume} {89}},\ \bibinfo {pages} {121117}
  (\bibinfo {year} {2014})}\BibitemShut {NoStop}%
\bibitem [{\citenamefont {Tan}\ \emph {et~al.}(2015)\citenamefont {Tan},
  \citenamefont {Hsu}, \citenamefont {Zeng}, \citenamefont {Hatnean},
  \citenamefont {Harrison}, \citenamefont {Zhu}, \citenamefont {Hartstein},
  \citenamefont {Kiourlappou}, \citenamefont {Srivastava}, \citenamefont
  {Johannes}, \citenamefont {Murphy}, \citenamefont {Park}, \citenamefont
  {Balicas}, \citenamefont {Lonzarich}, \citenamefont {Balakrishnan},\ and\
  \citenamefont {Sebastian}}]{Tan2015}%
  \BibitemOpen
  \bibfield  {author} {\bibinfo {author} {\bibfnamefont {B.~S.}\ \bibnamefont
  {Tan}}, \bibinfo {author} {\bibfnamefont {Y.-T.}\ \bibnamefont {Hsu}},
  \bibinfo {author} {\bibfnamefont {B.}~\bibnamefont {Zeng}}, \bibinfo {author}
  {\bibfnamefont {M.~Ciomaga}\ \bibnamefont {Hatnean}}, \bibinfo {author}
  {\bibfnamefont {N.}~\bibnamefont {Harrison}}, \bibinfo {author}
  {\bibfnamefont {Z.}~\bibnamefont {Zhu}}, \bibinfo {author} {\bibfnamefont
  {M.}~\bibnamefont {Hartstein}}, \bibinfo {author} {\bibfnamefont
  {M.}~\bibnamefont {Kiourlappou}}, \bibinfo {author} {\bibfnamefont
  {A.}~\bibnamefont {Srivastava}}, \bibinfo {author} {\bibfnamefont {M.~D.}\
  \bibnamefont {Johannes}}, \bibinfo {author} {\bibfnamefont {T.~P.}\
  \bibnamefont {Murphy}}, \bibinfo {author} {\bibfnamefont {J.-H.}\
  \bibnamefont {Park}}, \bibinfo {author} {\bibfnamefont {L.}~\bibnamefont
  {Balicas}}, \bibinfo {author} {\bibfnamefont {G.~G.}\ \bibnamefont
  {Lonzarich}}, \bibinfo {author} {\bibfnamefont {G.}~\bibnamefont
  {Balakrishnan}}, \ and\ \bibinfo {author} {\bibfnamefont {Suchitra~E.}\
  \bibnamefont {Sebastian}},\ }\bibfield  {title} {\enquote {\bibinfo {title}
  {Unconventional {F}ermi surface in an insulating state},}\ }\href {\doibase
  10.1126/science.aaa7974} {\bibfield  {journal} {\bibinfo  {journal}
  {Science}\ }\textbf {\bibinfo {volume} {349}},\ \bibinfo {pages} {287--290}
  (\bibinfo {year} {2015})}\BibitemShut {NoStop}%
\bibitem [{\citenamefont {Thomas}\ \emph {et~al.}(2018)\citenamefont {Thomas},
  \citenamefont {Ding}, \citenamefont {Ronning}, \citenamefont {Zapf},
  \citenamefont {Thompson}, \citenamefont {Fisk},\ and\ \citenamefont
  {Rosa}}]{Thomas2018}%
  \BibitemOpen
  \bibfield  {author} {\bibinfo {author} {\bibfnamefont {S.~M.}\ \bibnamefont
  {Thomas}}, \bibinfo {author} {\bibfnamefont {Xiaxin}\ \bibnamefont {Ding}},
  \bibinfo {author} {\bibfnamefont {F.}~\bibnamefont {Ronning}}, \bibinfo
  {author} {\bibfnamefont {V.}~\bibnamefont {Zapf}}, \bibinfo {author}
  {\bibfnamefont {J.~D.}\ \bibnamefont {Thompson}}, \bibinfo {author}
  {\bibfnamefont {Z.}~\bibnamefont {Fisk}}, \ and\ \bibinfo {author}
  {\bibfnamefont {P.F.S.}\ \bibnamefont {Rosa}},\ }\href@noop {} {\enquote
  {\bibinfo {title} {Quantum oscillations in flux-grown {SmB}$_6$ with embedded
  aluminium},}\ } (\bibinfo {year} {2018}),\ \bibinfo {note}
  {arXiv:1806.00117}\BibitemShut {NoStop}%
\bibitem [{\citenamefont {Allen}\ \emph {et~al.}(1980)\citenamefont {Allen},
  \citenamefont {Johansson}, \citenamefont {Lindau},\ and\ \citenamefont
  {Hagstrom}}]{Allen1980}%
  \BibitemOpen
  \bibfield  {author} {\bibinfo {author} {\bibfnamefont {J.~W.}\ \bibnamefont
  {Allen}}, \bibinfo {author} {\bibfnamefont {L.~I.}\ \bibnamefont
  {Johansson}}, \bibinfo {author} {\bibfnamefont {I.}~\bibnamefont {Lindau}}, \
  and\ \bibinfo {author} {\bibfnamefont {S.~B.}\ \bibnamefont {Hagstrom}},\
  }\bibfield  {title} {\enquote {\bibinfo {title} {Surface mixed valence in
  {S}m and {S}m{B}$_6$},}\ }\href {\doibase 10.1103/PhysRevB.21.1335}
  {\bibfield  {journal} {\bibinfo  {journal} {Phys. Rev. B}\ }\textbf {\bibinfo
  {volume} {21}},\ \bibinfo {pages} {1335--1343} (\bibinfo {year}
  {1980})}\BibitemShut {NoStop}%
\bibitem [{\citenamefont {Tarascon}\ \emph {et~al.}(1980)\citenamefont
  {Tarascon}, \citenamefont {Ishikawa}, \citenamefont {Chevalier},
  \citenamefont {Etourneau}, \citenamefont {Hagenmuller},\ and\ \citenamefont
  {M.}}]{Tarascon1980}%
  \BibitemOpen
  \bibfield  {author} {\bibinfo {author} {\bibfnamefont {J.~M.}\ \bibnamefont
  {Tarascon}}, \bibinfo {author} {\bibfnamefont {Y.}~\bibnamefont {Ishikawa}},
  \bibinfo {author} {\bibfnamefont {B.}~\bibnamefont {Chevalier}}, \bibinfo
  {author} {\bibfnamefont {J.}~\bibnamefont {Etourneau}}, \bibinfo {author}
  {\bibfnamefont {P.}~\bibnamefont {Hagenmuller}}, \ and\ \bibinfo {author}
  {\bibfnamefont {Kasaya}\ \bibnamefont {M.}},\ }\bibfield  {title} {\enquote
  {\bibinfo {title} {Temperature dependence of the samarium oxidation state in
  {S}m{B}$_6$ and {S}m$_{1-x}${L}a$_x${B}$_6$},}\ }\href {\doibase
  10.1051/jphys:0198000410100114100} {\bibfield  {journal} {\bibinfo  {journal}
  {J. Physique}\ }\textbf {\bibinfo {volume} {41}},\ \bibinfo {pages} {1141}
  (\bibinfo {year} {1980})}\BibitemShut {NoStop}%
\bibitem [{\citenamefont {Mizumaki}\ \emph {et~al.}(2009)\citenamefont
  {Mizumaki}, \citenamefont {Tsutsui},\ and\ \citenamefont
  {F.}}]{Mizumaki2009}%
  \BibitemOpen
  \bibfield  {author} {\bibinfo {author} {\bibfnamefont {M.}~\bibnamefont
  {Mizumaki}}, \bibinfo {author} {\bibfnamefont {S.}~\bibnamefont {Tsutsui}}, \
  and\ \bibinfo {author} {\bibfnamefont {Iga}\ \bibnamefont {F.}},\ }\bibfield
  {title} {\enquote {\bibinfo {title} {Temperature dependence of {S}m valence
  in {S}m{B}$_6$ studied by x-ray absorption spectroscopy},}\ }\href {\doibase
  doi:10.1088/1742-6596/176/1/012034} {\bibfield  {journal} {\bibinfo
  {journal} {J. Phys.: Conf. Ser.}\ }\textbf {\bibinfo {volume} {176}},\
  \bibinfo {pages} {012034} (\bibinfo {year} {2009})}\BibitemShut {NoStop}%
\bibitem [{\citenamefont {Butch}\ \emph {et~al.}(2016)\citenamefont {Butch},
  \citenamefont {Paglione}, \citenamefont {Chow}, \citenamefont {Xiao},
  \citenamefont {Marianetti}, \citenamefont {Booth},\ and\ \citenamefont
  {Jeffries}}]{Butch2016}%
  \BibitemOpen
  \bibfield  {author} {\bibinfo {author} {\bibfnamefont {N.~P.}\ \bibnamefont
  {Butch}}, \bibinfo {author} {\bibfnamefont {J.}~\bibnamefont {Paglione}},
  \bibinfo {author} {\bibfnamefont {P.}~\bibnamefont {Chow}}, \bibinfo {author}
  {\bibfnamefont {Y.}~\bibnamefont {Xiao}}, \bibinfo {author} {\bibfnamefont
  {C.~A.}\ \bibnamefont {Marianetti}}, \bibinfo {author} {\bibfnamefont
  {C.~H.}\ \bibnamefont {Booth}}, \ and\ \bibinfo {author} {\bibfnamefont
  {J.~R.}\ \bibnamefont {Jeffries}},\ }\bibfield  {title} {\enquote {\bibinfo
  {title} {Pressure-resistant intermediate valence in the {K}ondo insulator
  {S}m{B}$_6$},}\ }\href {\doibase 10.1103/PhysRevLett.116.156401} {\bibfield
  {journal} {\bibinfo  {journal} {Phys. Rev. Lett.}\ }\textbf {\bibinfo
  {volume} {116}},\ \bibinfo {pages} {156401} (\bibinfo {year}
  {2016})}\BibitemShut {NoStop}%
\bibitem [{\citenamefont {Utsumi}\ \emph {et~al.}(2017)\citenamefont {Utsumi},
  \citenamefont {Kasinathan}, \citenamefont {Ko}, \citenamefont {Agrestini},
  \citenamefont {Haverkort}, \citenamefont {Wirth}, \citenamefont {Wu},
  \citenamefont {Tsuei}, \citenamefont {Kim}, \citenamefont {Fisk},
  \citenamefont {Tanaka}, \citenamefont {Thalmeier},\ and\ \citenamefont
  {Tjeng}}]{Utsumi2017}%
  \BibitemOpen
  \bibfield  {author} {\bibinfo {author} {\bibfnamefont {Y.}~\bibnamefont
  {Utsumi}}, \bibinfo {author} {\bibfnamefont {D.}~\bibnamefont {Kasinathan}},
  \bibinfo {author} {\bibfnamefont {K.-T.}\ \bibnamefont {Ko}}, \bibinfo
  {author} {\bibfnamefont {S.}~\bibnamefont {Agrestini}}, \bibinfo {author}
  {\bibfnamefont {M.~W.}\ \bibnamefont {Haverkort}}, \bibinfo {author}
  {\bibfnamefont {S.}~\bibnamefont {Wirth}}, \bibinfo {author} {\bibfnamefont
  {Y.-H.}\ \bibnamefont {Wu}}, \bibinfo {author} {\bibfnamefont {K.-D.}\
  \bibnamefont {Tsuei}}, \bibinfo {author} {\bibfnamefont {D.-J.}\ \bibnamefont
  {Kim}}, \bibinfo {author} {\bibfnamefont {Z.}~\bibnamefont {Fisk}}, \bibinfo
  {author} {\bibfnamefont {A.}~\bibnamefont {Tanaka}}, \bibinfo {author}
  {\bibfnamefont {P.}~\bibnamefont {Thalmeier}}, \ and\ \bibinfo {author}
  {\bibfnamefont {L.~H.}\ \bibnamefont {Tjeng}},\ }\bibfield  {title} {\enquote
  {\bibinfo {title} {Bulk and surface electronic properties of {SmB}$_6$: A
  hard x-ray photoelectron spectroscopy study},}\ }\href {\doibase
  10.1103/PhysRevB.96.155130} {\bibfield  {journal} {\bibinfo  {journal} {Phys.
  Rev. B}\ }\textbf {\bibinfo {volume} {96}},\ \bibinfo {pages} {155130}
  (\bibinfo {year} {2017})}\BibitemShut {NoStop}%
\bibitem [{\citenamefont {Pellegrin}\ \emph {et~al.}(1993)\citenamefont
  {Pellegrin}, \citenamefont {N\"ucker}, \citenamefont {Fink}, \citenamefont
  {Molodtsov}, \citenamefont {Guti\'errez}, \citenamefont {Navas},
  \citenamefont {Strebel}, \citenamefont {Hu}, \citenamefont {Domke},
  \citenamefont {Kaindl}, \citenamefont {Uchida}, \citenamefont {Nakamura},
  \citenamefont {Markl}, \citenamefont {Klauda}, \citenamefont
  {Saemann-Ischenko}, \citenamefont {Krol}, \citenamefont {Peng}, \citenamefont
  {Li},\ and\ \citenamefont {Greene}}]{Pellegrin1993}%
  \BibitemOpen
  \bibfield  {author} {\bibinfo {author} {\bibfnamefont {E.}~\bibnamefont
  {Pellegrin}}, \bibinfo {author} {\bibfnamefont {N.}~\bibnamefont {N\"ucker}},
  \bibinfo {author} {\bibfnamefont {J.}~\bibnamefont {Fink}}, \bibinfo {author}
  {\bibfnamefont {S.~L.}\ \bibnamefont {Molodtsov}}, \bibinfo {author}
  {\bibfnamefont {A.}~\bibnamefont {Guti\'errez}}, \bibinfo {author}
  {\bibfnamefont {E.}~\bibnamefont {Navas}}, \bibinfo {author} {\bibfnamefont
  {O.}~\bibnamefont {Strebel}}, \bibinfo {author} {\bibfnamefont
  {Z.}~\bibnamefont {Hu}}, \bibinfo {author} {\bibfnamefont {M.}~\bibnamefont
  {Domke}}, \bibinfo {author} {\bibfnamefont {G.}~\bibnamefont {Kaindl}},
  \bibinfo {author} {\bibfnamefont {S.}~\bibnamefont {Uchida}}, \bibinfo
  {author} {\bibfnamefont {Y.}~\bibnamefont {Nakamura}}, \bibinfo {author}
  {\bibfnamefont {J.}~\bibnamefont {Markl}}, \bibinfo {author} {\bibfnamefont
  {M.}~\bibnamefont {Klauda}}, \bibinfo {author} {\bibfnamefont
  {G.}~\bibnamefont {Saemann-Ischenko}}, \bibinfo {author} {\bibfnamefont
  {A.}~\bibnamefont {Krol}}, \bibinfo {author} {\bibfnamefont {J.~L.}\
  \bibnamefont {Peng}}, \bibinfo {author} {\bibfnamefont {Z.~Y.}\ \bibnamefont
  {Li}}, \ and\ \bibinfo {author} {\bibfnamefont {R.~L.}\ \bibnamefont
  {Greene}},\ }\bibfield  {title} {\enquote {\bibinfo {title} {Orbital
  character of states at the {F}ermi level in {La}$_{2-x}${Sr}$_x${CuO}$_4$ and
  {R}$_{2-x}${Ce}$_x${CuO}$_4$ ({R=Nd,Sm})},}\ }\href {\doibase
  10.1103/PhysRevB.47.3354} {\bibfield  {journal} {\bibinfo  {journal} {Phys.
  Rev. B}\ }\textbf {\bibinfo {volume} {47}},\ \bibinfo {pages} {3354--3367}
  (\bibinfo {year} {1993})}\BibitemShut {NoStop}%
\bibitem [{\citenamefont {Lea}\ \emph {et~al.}(1962)\citenamefont {Lea},
  \citenamefont {Leask},\ and\ \citenamefont {Wolf}}]{Lea1962}%
  \BibitemOpen
  \bibfield  {author} {\bibinfo {author} {\bibfnamefont {K.~R.}\ \bibnamefont
  {Lea}}, \bibinfo {author} {\bibfnamefont {M.~J.~M.}\ \bibnamefont {Leask}}, \
  and\ \bibinfo {author} {\bibfnamefont {W.~P.}\ \bibnamefont {Wolf}},\
  }\bibfield  {title} {\enquote {\bibinfo {title} {The raising of angular
  momentum degeneracy of f-electron terms by cubic crystal fields},}\ }\href
  {\doibase http://dx.doi.org/10.1016/0022-3697(62)90192-0} {\bibfield
  {journal} {\bibinfo  {journal} {J. Phys. Chem. Solids}\ }\textbf {\bibinfo
  {volume} {23}},\ \bibinfo {pages} {1381 -- 1405} (\bibinfo {year}
  {1962})}\BibitemShut {NoStop}%
\bibitem [{\citenamefont {Yanase}\ and\ \citenamefont
  {Harima}(1992)}]{Yanase1992}%
  \BibitemOpen
  \bibfield  {author} {\bibinfo {author} {\bibfnamefont {A.}~\bibnamefont
  {Yanase}}\ and\ \bibinfo {author} {\bibfnamefont {H.}~\bibnamefont
  {Harima}},\ }\bibfield  {title} {\enquote {\bibinfo {title} {Band
  calculations on {YbB}$_{12}$, {SmB}$_6$ and {CeNiSn}},}\ }\href {\doibase
  10.1143/PTPS.108.19} {\bibfield  {journal} {\bibinfo  {journal} {Prog. Theo.
  Phys. Supp.}\ }\textbf {\bibinfo {volume} {108}},\ \bibinfo {pages} {19--25}
  (\bibinfo {year} {1992})}\BibitemShut {NoStop}%
\bibitem [{\citenamefont {Antonov}\ \emph {et~al.}(2002)\citenamefont
  {Antonov}, \citenamefont {Harmon},\ and\ \citenamefont
  {Yaresko}}]{Antonov2002}%
  \BibitemOpen
  \bibfield  {author} {\bibinfo {author} {\bibfnamefont {V.~N.}\ \bibnamefont
  {Antonov}}, \bibinfo {author} {\bibfnamefont {B.~N.}\ \bibnamefont {Harmon}},
  \ and\ \bibinfo {author} {\bibfnamefont {A.~N.}\ \bibnamefont {Yaresko}},\
  }\bibfield  {title} {\enquote {\bibinfo {title} {Electronic structure of
  mixed-valence semiconductors in the {LSDA}+{$U$} approximation. {II}.
  {SmB}$_6$ and {YbB}$_{12}$},}\ }\href {\doibase 10.1103/PhysRevB.66.165209}
  {\bibfield  {journal} {\bibinfo  {journal} {Phys. Rev. B}\ }\textbf {\bibinfo
  {volume} {66}},\ \bibinfo {pages} {165209} (\bibinfo {year}
  {2002})}\BibitemShut {NoStop}%
\bibitem [{\citenamefont {Kang}\ \emph {et~al.}(2015)\citenamefont {Kang},
  \citenamefont {Kim}, \citenamefont {Kim}, \citenamefont {Kang}, \citenamefont
  {Denlinger},\ and\ \citenamefont {Min}}]{Kang2015}%
  \BibitemOpen
  \bibfield  {author} {\bibinfo {author} {\bibfnamefont {C.~J.}\ \bibnamefont
  {Kang}}, \bibinfo {author} {\bibfnamefont {J.}~\bibnamefont {Kim}}, \bibinfo
  {author} {\bibfnamefont {K.}~\bibnamefont {Kim}}, \bibinfo {author}
  {\bibfnamefont {J.}~\bibnamefont {Kang}}, \bibinfo {author} {\bibfnamefont
  {J.~D.}\ \bibnamefont {Denlinger}}, \ and\ \bibinfo {author} {\bibfnamefont
  {B.~Il}\ \bibnamefont {Min}},\ }\bibfield  {title} {\enquote {\bibinfo
  {title} {Band symmetries of mixed-valence topological insulator:
  {SmB}$_6$},}\ }\href {\doibase 10.7566/JPSJ.84.024722} {\bibfield  {journal}
  {\bibinfo  {journal} {J. Phys. Soc. Jpn.}\ }\textbf {\bibinfo {volume}
  {84}},\ \bibinfo {pages} {024722} (\bibinfo {year} {2015})}\BibitemShut
  {NoStop}%
\bibitem [{\citenamefont {Singh}\ and\ \citenamefont {Lee}(2018)}]{Singh2018}%
  \BibitemOpen
  \bibfield  {author} {\bibinfo {author} {\bibfnamefont {C.~N.}\ \bibnamefont
  {Singh}}\ and\ \bibinfo {author} {\bibfnamefont {W.~C.}\ \bibnamefont
  {Lee}},\ }\bibfield  {title} {\enquote {\bibinfo {title} {Importance of
  orbital fluctuations for the magnetic dynamics in the heavy-fermion compound
  {SmB}$_6$},}\ }\href {\doibase 10.1103/PhysRevB.97.241107} {\bibfield
  {journal} {\bibinfo  {journal} {Phys. Rev. B}\ }\textbf {\bibinfo {volume}
  {97}},\ \bibinfo {pages} {241107} (\bibinfo {year} {2018})}\BibitemShut
  {NoStop}%
\bibitem [{\citenamefont {Sundermann}\ \emph {et~al.}(2018)\citenamefont
  {Sundermann}, \citenamefont {Yavas}, \citenamefont {Chen}, \citenamefont
  {Kim}, \citenamefont {Fisk}, \citenamefont {Kasinathan}, \citenamefont
  {Haverkort}, \citenamefont {Thalmeier}, \citenamefont {Severing},\ and\
  \citenamefont {Tjeng}}]{Sundermann2018}%
  \BibitemOpen
  \bibfield  {author} {\bibinfo {author} {\bibfnamefont {M.}~\bibnamefont
  {Sundermann}}, \bibinfo {author} {\bibfnamefont {H.}~\bibnamefont {Yavas}},
  \bibinfo {author} {\bibfnamefont {K.}~\bibnamefont {Chen}}, \bibinfo {author}
  {\bibfnamefont {D.~J.}\ \bibnamefont {Kim}}, \bibinfo {author} {\bibfnamefont
  {Z.}~\bibnamefont {Fisk}}, \bibinfo {author} {\bibfnamefont {D.}~\bibnamefont
  {Kasinathan}}, \bibinfo {author} {\bibfnamefont {M.~W.}\ \bibnamefont
  {Haverkort}}, \bibinfo {author} {\bibfnamefont {P.}~\bibnamefont
  {Thalmeier}}, \bibinfo {author} {\bibfnamefont {A.}~\bibnamefont {Severing}},
  \ and\ \bibinfo {author} {\bibfnamefont {L.~H.}\ \bibnamefont {Tjeng}},\
  }\bibfield  {title} {\enquote {\bibinfo {title} {$4f$ crystal field ground
  state of the strongly correlated topological insulator {SmB}$_6$},}\ }\href
  {\doibase 10.1103/PhysRevLett.120.016402} {\bibfield  {journal} {\bibinfo
  {journal} {Phys. Rev. Lett.}\ }\textbf {\bibinfo {volume} {120}},\ \bibinfo
  {pages} {016402} (\bibinfo {year} {2018})}\BibitemShut {NoStop}%
\bibitem [{foo({\natexlab{a}})}]{footnote1}%
  \BibitemOpen
  \href@noop {} {} ({\natexlab{a}}),\ \bibinfo {note} {{N}on-resonant inelastic
  x-ray scattering ({NIXS}) specifically targets the CF ground state symmetry,
  not the CF splitting.}\BibitemShut {Stop}%
\bibitem [{\citenamefont {Loewenhaupt}\ and\ \citenamefont
  {Prager}(1986)}]{Loewenhaupt1986}%
  \BibitemOpen
  \bibfield  {author} {\bibinfo {author} {\bibfnamefont {M.}~\bibnamefont
  {Loewenhaupt}}\ and\ \bibinfo {author} {\bibfnamefont {M.}~\bibnamefont
  {Prager}},\ }\bibfield  {title} {\enquote {\bibinfo {title} {Crystal fields
  in {PrB}$_6$ and {NdB}$_6$},}\ }\href {\doibase 10.1007/BF01323430}
  {\bibfield  {journal} {\bibinfo  {journal} {Z. Phys. B Cond. Mat.}\ }\textbf
  {\bibinfo {volume} {62}},\ \bibinfo {pages} {195--199} (\bibinfo {year}
  {1986})}\BibitemShut {NoStop}%
\bibitem [{\citenamefont {Frick}\ and\ \citenamefont
  {Loewenhaupt}(1986)}]{Frick1986}%
  \BibitemOpen
  \bibfield  {author} {\bibinfo {author} {\bibfnamefont {B.}~\bibnamefont
  {Frick}}\ and\ \bibinfo {author} {\bibfnamefont {M.}~\bibnamefont
  {Loewenhaupt}},\ }\bibfield  {title} {\enquote {\bibinfo {title} {Crystal
  field spectroscopy by inelastic neutron scattering},}\ }\href {\doibase
  10.1007/BF01309242} {\bibfield  {journal} {\bibinfo  {journal} {Z. Phys. B
  Cond. Mat.}\ }\textbf {\bibinfo {volume} {63}},\ \bibinfo {pages} {213--230}
  (\bibinfo {year} {1986})}\BibitemShut {NoStop}%
\bibitem [{\citenamefont {Kim}\ \emph {et~al.}(2014{\natexlab{b}})\citenamefont
  {Kim}, \citenamefont {Kim}, \citenamefont {Kang}, \citenamefont {Kim},
  \citenamefont {Choi}, \citenamefont {Kang}, \citenamefont {Denlinger},\ and\
  \citenamefont {Min}}]{Kim2014a}%
  \BibitemOpen
  \bibfield  {author} {\bibinfo {author} {\bibfnamefont {J.}~\bibnamefont
  {Kim}}, \bibinfo {author} {\bibfnamefont {K.}~\bibnamefont {Kim}}, \bibinfo
  {author} {\bibfnamefont {C.-J.}\ \bibnamefont {Kang}}, \bibinfo {author}
  {\bibfnamefont {S.}~\bibnamefont {Kim}}, \bibinfo {author} {\bibfnamefont
  {H.~C.}\ \bibnamefont {Choi}}, \bibinfo {author} {\bibfnamefont {J.-S.}\
  \bibnamefont {Kang}}, \bibinfo {author} {\bibfnamefont {J.~D.}\ \bibnamefont
  {Denlinger}}, \ and\ \bibinfo {author} {\bibfnamefont {B.~I.}\ \bibnamefont
  {Min}},\ }\bibfield  {title} {\enquote {\bibinfo {title}
  {Termination-dependent surface in-gap states in a potential mixed-valent
  topological insulator: {SmB}$_6$},}\ }\href {\doibase
  10.1103/PhysRevB.90.075131} {\bibfield  {journal} {\bibinfo  {journal} {Phys.
  Rev. B}\ }\textbf {\bibinfo {volume} {90}},\ \bibinfo {pages} {075131}
  (\bibinfo {year} {2014}{\natexlab{b}})}\BibitemShut {NoStop}%
\bibitem [{\citenamefont {Alekseev}\ \emph {et~al.}(1993)\citenamefont
  {Alekseev}, \citenamefont {Lazukov}, \citenamefont {Osborn}, \citenamefont
  {Rainford}, \citenamefont {Sadikov}, \citenamefont {Konovalova},\ and\
  \citenamefont {Paderno}}]{Alekseev1993}%
  \BibitemOpen
  \bibfield  {author} {\bibinfo {author} {\bibfnamefont {P.~A.}\ \bibnamefont
  {Alekseev}}, \bibinfo {author} {\bibfnamefont {V.~N.}\ \bibnamefont
  {Lazukov}}, \bibinfo {author} {\bibfnamefont {R.}~\bibnamefont {Osborn}},
  \bibinfo {author} {\bibfnamefont {B.~D.}\ \bibnamefont {Rainford}}, \bibinfo
  {author} {\bibfnamefont {I.~P.}\ \bibnamefont {Sadikov}}, \bibinfo {author}
  {\bibfnamefont {E.~S.}\ \bibnamefont {Konovalova}}, \ and\ \bibinfo {author}
  {\bibfnamefont {Yu.~B.}\ \bibnamefont {Paderno}},\ }\bibfield  {title}
  {\enquote {\bibinfo {title} {Neutron scattering study of the
  intermediate-valent ground state in {SmB}$_6$},}\ }\href
  {http://stacks.iop.org/0295-5075/23/i=5/a=008} {\bibfield  {journal}
  {\bibinfo  {journal} {Euro. Phys. Lett.}\ }\textbf {\bibinfo {volume} {23}},\
  \bibinfo {pages} {347} (\bibinfo {year} {1993})}\BibitemShut {NoStop}%
\bibitem [{\citenamefont {Alekseev}\ \emph {et~al.}(1995)\citenamefont
  {Alekseev}, \citenamefont {Mignot}, \citenamefont {Rossat-Mignod},
  \citenamefont {Lazukov}, \citenamefont {Sadikov}, \citenamefont
  {Konovalova},\ and\ \citenamefont {Paderno}}]{Alekseev1995}%
  \BibitemOpen
  \bibfield  {author} {\bibinfo {author} {\bibfnamefont {P.~A.}\ \bibnamefont
  {Alekseev}}, \bibinfo {author} {\bibfnamefont {J.~M.}\ \bibnamefont
  {Mignot}}, \bibinfo {author} {\bibfnamefont {J.}~\bibnamefont
  {Rossat-Mignod}}, \bibinfo {author} {\bibfnamefont {V.~N.}\ \bibnamefont
  {Lazukov}}, \bibinfo {author} {\bibfnamefont {I.~P.}\ \bibnamefont
  {Sadikov}}, \bibinfo {author} {\bibfnamefont {E.~S.}\ \bibnamefont
  {Konovalova}}, \ and\ \bibinfo {author} {\bibfnamefont {Yu~B.}\ \bibnamefont
  {Paderno}},\ }\bibfield  {title} {\enquote {\bibinfo {title} {Magnetic
  excitation spectrum of mixed-valence {S}m{B}$_6$ studied by neutron
  scattering on a single crystal},}\ }\href
  {http://stacks.iop.org/0953-8984/7/i=2/a=007} {\bibfield  {journal} {\bibinfo
   {journal} {J. Physics: Cond. Matt.}\ }\textbf {\bibinfo {volume} {7}},\
  \bibinfo {pages} {289} (\bibinfo {year} {1995})}\BibitemShut {NoStop}%
\bibitem [{\citenamefont {Fuhrman}\ \emph {et~al.}(2015)\citenamefont
  {Fuhrman}, \citenamefont {Leiner}, \citenamefont
  {Nikoli\ifmmode~\acute{c}\else \'{c}\fi{}}, \citenamefont {Granroth},
  \citenamefont {Stone}, \citenamefont {Lumsden}, \citenamefont
  {DeBeer-Schmitt}, \citenamefont {Alekseev}, \citenamefont {Mignot},
  \citenamefont {Koohpayeh}, \citenamefont {Cottingham}, \citenamefont
  {Phelan}, \citenamefont {Schoop}, \citenamefont {McQueen},\ and\
  \citenamefont {Broholm}}]{Fuhrman2015}%
  \BibitemOpen
  \bibfield  {author} {\bibinfo {author} {\bibfnamefont {W.~T.}\ \bibnamefont
  {Fuhrman}}, \bibinfo {author} {\bibfnamefont {J.}~\bibnamefont {Leiner}},
  \bibinfo {author} {\bibfnamefont {P.}~\bibnamefont
  {Nikoli\ifmmode~\acute{c}\else \'{c}\fi{}}}, \bibinfo {author} {\bibfnamefont
  {G.~E.}\ \bibnamefont {Granroth}}, \bibinfo {author} {\bibfnamefont {M.~B.}\
  \bibnamefont {Stone}}, \bibinfo {author} {\bibfnamefont {M.~D.}\ \bibnamefont
  {Lumsden}}, \bibinfo {author} {\bibfnamefont {L.}~\bibnamefont
  {DeBeer-Schmitt}}, \bibinfo {author} {\bibfnamefont {P.~A.}\ \bibnamefont
  {Alekseev}}, \bibinfo {author} {\bibfnamefont {J.-M.}\ \bibnamefont
  {Mignot}}, \bibinfo {author} {\bibfnamefont {S.~M.}\ \bibnamefont
  {Koohpayeh}}, \bibinfo {author} {\bibfnamefont {P.}~\bibnamefont
  {Cottingham}}, \bibinfo {author} {\bibfnamefont {W.~A.}\ \bibnamefont
  {Phelan}}, \bibinfo {author} {\bibfnamefont {L.}~\bibnamefont {Schoop}},
  \bibinfo {author} {\bibfnamefont {T.~M.}\ \bibnamefont {McQueen}}, \ and\
  \bibinfo {author} {\bibfnamefont {C.}~\bibnamefont {Broholm}},\ }\bibfield
  {title} {\enquote {\bibinfo {title} {Interaction driven subgap spin exciton
  in the {K}ondo insulator {SmB}$_6$},}\ }\href {\doibase
  10.1103/PhysRevLett.114.036401} {\bibfield  {journal} {\bibinfo  {journal}
  {Phys. Rev. Lett.}\ }\textbf {\bibinfo {volume} {114}},\ \bibinfo {pages}
  {036401} (\bibinfo {year} {2015})}\BibitemShut {NoStop}%
\bibitem [{\citenamefont {Fuhrman}\ \emph {et~al.}(2018)\citenamefont
  {Fuhrman}, \citenamefont {Chamorro}, \citenamefont {Alekseev}, \citenamefont
  {Mignot}, \citenamefont {Keller}, \citenamefont {Rodriguez-Rivera},
  \citenamefont {Qiu}, \citenamefont {Nikolić}, \citenamefont {McQueen},\ and\
  \citenamefont {Broholm}}]{Fuhrman2018}%
  \BibitemOpen
  \bibfield  {author} {\bibinfo {author} {\bibfnamefont {W.~T.}\ \bibnamefont
  {Fuhrman}}, \bibinfo {author} {\bibfnamefont {J.~R.}\ \bibnamefont
  {Chamorro}}, \bibinfo {author} {\bibfnamefont {P.~A.}\ \bibnamefont
  {Alekseev}}, \bibinfo {author} {\bibfnamefont {J.-M.}\ \bibnamefont
  {Mignot}}, \bibinfo {author} {\bibfnamefont {T.}~\bibnamefont {Keller}},
  \bibinfo {author} {\bibfnamefont {J.~A.}\ \bibnamefont {Rodriguez-Rivera}},
  \bibinfo {author} {\bibfnamefont {Y.}~\bibnamefont {Qiu}}, \bibinfo {author}
  {\bibfnamefont {P.}~\bibnamefont {Nikolić}}, \bibinfo {author}
  {\bibfnamefont {T.~M.}\ \bibnamefont {McQueen}}, \ and\ \bibinfo {author}
  {\bibfnamefont {C.~L.}\ \bibnamefont {Broholm}},\ }\bibfield  {title}
  {\enquote {\bibinfo {title} {Screened moments and extrinsic in-gap states in
  samarium hexaboride},}\ }\href {\doibase 10.1038/s41467-018-04007-z}
  {\bibfield  {journal} {\bibinfo  {journal} {Nature Comm.}\ }\textbf {\bibinfo
  {volume} {9}},\ \bibinfo {pages} {1539} (\bibinfo {year} {2018})}\BibitemShut
  {NoStop}%
\bibitem [{\citenamefont {Alekseev}\ \emph {et~al.}(2016)\citenamefont
  {Alekseev}, \citenamefont {Mignot}, \citenamefont {Savchenkov},\ and\
  \citenamefont {Lazukov}}]{Alekseev2016}%
  \BibitemOpen
  \bibfield  {author} {\bibinfo {author} {\bibfnamefont {P.~A.}\ \bibnamefont
  {Alekseev}}, \bibinfo {author} {\bibfnamefont {J.~M.}\ \bibnamefont
  {Mignot}}, \bibinfo {author} {\bibfnamefont {P.~S.}\ \bibnamefont
  {Savchenkov}}, \ and\ \bibinfo {author} {\bibfnamefont {V.~N.}\ \bibnamefont
  {Lazukov}},\ }\bibfield  {title} {\enquote {\bibinfo {title} {First evidence
  for a {S}m$^{3+}$-type contribution to the magnetic form factor in the
  quasielastic spectral response of intermediate valence {SmB}$_6$},}\ }\href
  {\doibase 10.1134/S0021364016100015} {\bibfield  {journal} {\bibinfo
  {journal} {J. Exp. and Theo. Phys. Lett.}\ }\textbf {\bibinfo {volume}
  {103}},\ \bibinfo {pages} {636--642} (\bibinfo {year} {2016})}\BibitemShut
  {NoStop}%
\bibitem [{\citenamefont {Vyalikh}\ \emph {et~al.}(2010)\citenamefont
  {Vyalikh}, \citenamefont {Danzenb\"acher}, \citenamefont {Kucherenko},
  \citenamefont {Kummer}, \citenamefont {Krellner}, \citenamefont {Geibel},
  \citenamefont {Holder}, \citenamefont {Kim}, \citenamefont {Laubschat},
  \citenamefont {Shi}, \citenamefont {Patthey}, \citenamefont {Follath},\ and\
  \citenamefont {Molodtsov}}]{Vyalikh2010}%
  \BibitemOpen
  \bibfield  {author} {\bibinfo {author} {\bibfnamefont {D.~V.}\ \bibnamefont
  {Vyalikh}}, \bibinfo {author} {\bibfnamefont {S.}~\bibnamefont
  {Danzenb\"acher}}, \bibinfo {author} {\bibfnamefont {Yu.}\ \bibnamefont
  {Kucherenko}}, \bibinfo {author} {\bibfnamefont {K.}~\bibnamefont {Kummer}},
  \bibinfo {author} {\bibfnamefont {C.}~\bibnamefont {Krellner}}, \bibinfo
  {author} {\bibfnamefont {C.}~\bibnamefont {Geibel}}, \bibinfo {author}
  {\bibfnamefont {M.~G.}\ \bibnamefont {Holder}}, \bibinfo {author}
  {\bibfnamefont {T.~K.}\ \bibnamefont {Kim}}, \bibinfo {author} {\bibfnamefont
  {C.}~\bibnamefont {Laubschat}}, \bibinfo {author} {\bibfnamefont
  {M.}~\bibnamefont {Shi}}, \bibinfo {author} {\bibfnamefont {L.}~\bibnamefont
  {Patthey}}, \bibinfo {author} {\bibfnamefont {R.}~\bibnamefont {Follath}}, \
  and\ \bibinfo {author} {\bibfnamefont {S.~L.}\ \bibnamefont {Molodtsov}},\
  }\bibfield  {title} {\enquote {\bibinfo {title} {$k$ dependence of the
  crystal-field splittings of $4f$ states in rare-earth systems},}\ }\href
  {\doibase 10.1103/PhysRevLett.105.237601} {\bibfield  {journal} {\bibinfo
  {journal} {Phys. Rev. Lett.}\ }\textbf {\bibinfo {volume} {105}},\ \bibinfo
  {pages} {237601} (\bibinfo {year} {2010})}\BibitemShut {NoStop}%
\bibitem [{\citenamefont {Amorese}\ \emph
  {et~al.}(2018{\natexlab{a}})\citenamefont {Amorese}, \citenamefont
  {Caroca-Canales}, \citenamefont {Seiro}, \citenamefont {Krellner},
  \citenamefont {Ghiringhelli}, \citenamefont {Brookes}, \citenamefont
  {Vyalikh}, \citenamefont {Geibel},\ and\ \citenamefont {Kummer}}]{Amorese_a}%
  \BibitemOpen
  \bibfield  {author} {\bibinfo {author} {\bibfnamefont {A.}~\bibnamefont
  {Amorese}}, \bibinfo {author} {\bibfnamefont {N.}~\bibnamefont
  {Caroca-Canales}}, \bibinfo {author} {\bibfnamefont {S.}~\bibnamefont
  {Seiro}}, \bibinfo {author} {\bibfnamefont {C.}~\bibnamefont {Krellner}},
  \bibinfo {author} {\bibfnamefont {G.}~\bibnamefont {Ghiringhelli}}, \bibinfo
  {author} {\bibfnamefont {N.~B.}\ \bibnamefont {Brookes}}, \bibinfo {author}
  {\bibfnamefont {D.~V.}\ \bibnamefont {Vyalikh}}, \bibinfo {author}
  {\bibfnamefont {C.}~\bibnamefont {Geibel}}, \ and\ \bibinfo {author}
  {\bibfnamefont {K.}~\bibnamefont {Kummer}},\ }\bibfield  {title} {\enquote
  {\bibinfo {title} {Crystal electric field in {CeRh}$_2${Si}$_2$ studied with
  high-resolution resonant inelastic soft x-ray scattering},}\ }\href {\doibase
  10.1103/PhysRevB.97.245130} {\bibfield  {journal} {\bibinfo  {journal} {Phys.
  Rev. B}\ }\textbf {\bibinfo {volume} {97}},\ \bibinfo {pages} {245130}
  (\bibinfo {year} {2018}{\natexlab{a}})}\BibitemShut {NoStop}%
\bibitem [{\citenamefont {Amorese}\ \emph
  {et~al.}(2018{\natexlab{b}})\citenamefont {Amorese}, \citenamefont {Kummer},
  \citenamefont {Brookes}, \citenamefont {Stockert}, \citenamefont {Adroja},
  \citenamefont {Strydom}, \citenamefont {Sidorenko}, \citenamefont {Winkler},
  \citenamefont {Zocco}, \citenamefont {Prokofiev}, \citenamefont {Paschen},
  \citenamefont {Haverkort}, \citenamefont {Tjeng},\ and\ \citenamefont
  {Severing}}]{Amorese_b}%
  \BibitemOpen
  \bibfield  {author} {\bibinfo {author} {\bibfnamefont {A.}~\bibnamefont
  {Amorese}}, \bibinfo {author} {\bibfnamefont {K.}~\bibnamefont {Kummer}},
  \bibinfo {author} {\bibfnamefont {N.~B.}\ \bibnamefont {Brookes}}, \bibinfo
  {author} {\bibfnamefont {O.}~\bibnamefont {Stockert}}, \bibinfo {author}
  {\bibfnamefont {D.~T.}\ \bibnamefont {Adroja}}, \bibinfo {author}
  {\bibfnamefont {A.~M.}\ \bibnamefont {Strydom}}, \bibinfo {author}
  {\bibfnamefont {A.}~\bibnamefont {Sidorenko}}, \bibinfo {author}
  {\bibfnamefont {H.}~\bibnamefont {Winkler}}, \bibinfo {author} {\bibfnamefont
  {D.~A.}\ \bibnamefont {Zocco}}, \bibinfo {author} {\bibfnamefont
  {A.}~\bibnamefont {Prokofiev}}, \bibinfo {author} {\bibfnamefont
  {S.}~\bibnamefont {Paschen}}, \bibinfo {author} {\bibfnamefont {M.~W.}\
  \bibnamefont {Haverkort}}, \bibinfo {author} {\bibfnamefont {L.~H.}\
  \bibnamefont {Tjeng}}, \ and\ \bibinfo {author} {\bibfnamefont
  {A.}~\bibnamefont {Severing}},\ }\bibfield  {title} {\enquote {\bibinfo
  {title} {Determining the local low-energy excitations in the {K}ondo
  semimetal {CeRu}$_4${Sn}$_6$ using resonant inelastic x-ray scattering},}\
  }\href {\doibase 10.1103/PhysRevB.98.081116} {\bibfield  {journal} {\bibinfo
  {journal} {Phys. Rev. B}\ }\textbf {\bibinfo {volume} {98}},\ \bibinfo
  {pages} {081116} (\bibinfo {year} {2018}{\natexlab{b}})}\BibitemShut
  {NoStop}%
\bibitem [{\citenamefont {Kotani}\ and\ \citenamefont
  {Shin}(2001)}]{Kotani2001}%
  \BibitemOpen
  \bibfield  {author} {\bibinfo {author} {\bibfnamefont {A.}~\bibnamefont
  {Kotani}}\ and\ \bibinfo {author} {\bibfnamefont {S.}~\bibnamefont {Shin}},\
  }\bibfield  {title} {\enquote {\bibinfo {title} {Resonant inelastic x-ray
  scattering spectra for electrons in solids},}\ }\href {\doibase
  10.1103/RevModPhys.73.203} {\bibfield  {journal} {\bibinfo  {journal} {Rev.
  Mod. Phys.}\ }\textbf {\bibinfo {volume} {73}},\ \bibinfo {pages} {203--246}
  (\bibinfo {year} {2001})}\BibitemShut {NoStop}%
\bibitem [{\citenamefont {Dallera}\ \emph {et~al.}(2002)\citenamefont
  {Dallera}, \citenamefont {Grioni}, \citenamefont {Shukla}, \citenamefont
  {Vank\'o}, \citenamefont {Sarrao}, \citenamefont {Rueff},\ and\ \citenamefont
  {Cox}}]{Dallera2002}%
  \BibitemOpen
  \bibfield  {author} {\bibinfo {author} {\bibfnamefont {C.}~\bibnamefont
  {Dallera}}, \bibinfo {author} {\bibfnamefont {M.}~\bibnamefont {Grioni}},
  \bibinfo {author} {\bibfnamefont {A.}~\bibnamefont {Shukla}}, \bibinfo
  {author} {\bibfnamefont {G.}~\bibnamefont {Vank\'o}}, \bibinfo {author}
  {\bibfnamefont {J.~L.}\ \bibnamefont {Sarrao}}, \bibinfo {author}
  {\bibfnamefont {J.~P.}\ \bibnamefont {Rueff}}, \ and\ \bibinfo {author}
  {\bibfnamefont {D.~L.}\ \bibnamefont {Cox}},\ }\bibfield  {title} {\enquote
  {\bibinfo {title} {New spectroscopy solves an old puzzle: The {K}ondo scale
  in heavy fermions},}\ }\href {\doibase 10.1103/PhysRevLett.88.196403}
  {\bibfield  {journal} {\bibinfo  {journal} {Phys. Rev. Lett.}\ }\textbf
  {\bibinfo {volume} {88}},\ \bibinfo {pages} {196403} (\bibinfo {year}
  {2002})}\BibitemShut {NoStop}%
\bibitem [{\citenamefont {Haak}\ \emph {et~al.}(1978)\citenamefont {Haak},
  \citenamefont {Sawatzky},\ and\ \citenamefont {Thomas}}]{Haak1978}%
  \BibitemOpen
  \bibfield  {author} {\bibinfo {author} {\bibfnamefont {H.~W.}\ \bibnamefont
  {Haak}}, \bibinfo {author} {\bibfnamefont {G.~A.}\ \bibnamefont {Sawatzky}},
  \ and\ \bibinfo {author} {\bibfnamefont {T.~D.}\ \bibnamefont {Thomas}},\
  }\bibfield  {title} {\enquote {\bibinfo {title} {Auger-photoelectron
  coincidence measurements in copper},}\ }\href {\doibase
  10.1103/PhysRevLett.41.1825} {\bibfield  {journal} {\bibinfo  {journal}
  {Phys. Rev. Lett.}\ }\textbf {\bibinfo {volume} {41}},\ \bibinfo {pages}
  {1825--1827} (\bibinfo {year} {1978})}\BibitemShut {NoStop}%
\bibitem [{\citenamefont {Jensen}\ \emph {et~al.}(1989)\citenamefont {Jensen},
  \citenamefont {Bartynski}, \citenamefont {Hulbert}, \citenamefont {Johnson},\
  and\ \citenamefont {Garrett}}]{Jensen1989}%
  \BibitemOpen
  \bibfield  {author} {\bibinfo {author} {\bibfnamefont {E.}~\bibnamefont
  {Jensen}}, \bibinfo {author} {\bibfnamefont {R.~A.}\ \bibnamefont
  {Bartynski}}, \bibinfo {author} {\bibfnamefont {S.~L.}\ \bibnamefont
  {Hulbert}}, \bibinfo {author} {\bibfnamefont {E.~D.}\ \bibnamefont
  {Johnson}}, \ and\ \bibinfo {author} {\bibfnamefont {R.}~\bibnamefont
  {Garrett}},\ }\bibfield  {title} {\enquote {\bibinfo {title} {Line narrowing
  in photoemission by coincidence spectroscopy},}\ }\href {\doibase
  10.1103/PhysRevLett.62.71} {\bibfield  {journal} {\bibinfo  {journal} {Phys.
  Rev. Lett.}\ }\textbf {\bibinfo {volume} {62}},\ \bibinfo {pages} {71--73}
  (\bibinfo {year} {1989})}\BibitemShut {NoStop}%
\bibitem [{\citenamefont {H\"am\"al\"ainen}\ \emph {et~al.}(1991)\citenamefont
  {H\"am\"al\"ainen}, \citenamefont {Siddons}, \citenamefont {Hastings},\ and\
  \citenamefont {Berman}}]{Berman1991}%
  \BibitemOpen
  \bibfield  {author} {\bibinfo {author} {\bibfnamefont {K.}~\bibnamefont
  {H\"am\"al\"ainen}}, \bibinfo {author} {\bibfnamefont {D.~P.}\ \bibnamefont
  {Siddons}}, \bibinfo {author} {\bibfnamefont {J.~B.}\ \bibnamefont
  {Hastings}}, \ and\ \bibinfo {author} {\bibfnamefont {L.~E.}\ \bibnamefont
  {Berman}},\ }\bibfield  {title} {\enquote {\bibinfo {title} {Elimination of
  the inner-shell lifetime broadening in x-ray-absorption spectroscopy},}\
  }\href {\doibase 10.1103/PhysRevLett.67.2850} {\bibfield  {journal} {\bibinfo
   {journal} {Phys. Rev. Lett.}\ }\textbf {\bibinfo {volume} {67}},\ \bibinfo
  {pages} {2850--2853} (\bibinfo {year} {1991})}\BibitemShut {NoStop}%
\bibitem [{\citenamefont {Carra}\ \emph {et~al.}(1995)\citenamefont {Carra},
  \citenamefont {Fabrizio},\ and\ \citenamefont {Thole}}]{Carra1995}%
  \BibitemOpen
  \bibfield  {author} {\bibinfo {author} {\bibfnamefont {P.}~\bibnamefont
  {Carra}}, \bibinfo {author} {\bibfnamefont {M.}~\bibnamefont {Fabrizio}}, \
  and\ \bibinfo {author} {\bibfnamefont {B.~T.}\ \bibnamefont {Thole}},\
  }\bibfield  {title} {\enquote {\bibinfo {title} {High resolution x-ray
  resonant raman scattering},}\ }\href {\doibase 10.1103/PhysRevLett.74.3700}
  {\bibfield  {journal} {\bibinfo  {journal} {Phys. Rev. Lett.}\ }\textbf
  {\bibinfo {volume} {74}},\ \bibinfo {pages} {3700--3703} (\bibinfo {year}
  {1995})}\BibitemShut {NoStop}%
\bibitem [{foo({\natexlab{b}})}]{footnote2}%
  \BibitemOpen
  \href@noop {} {} ({\natexlab{b}}),\ \bibinfo {note} {{H}ere $\check
  A_k^m=A_k^m\langle r^k\rangle$, with $\langle r^k\rangle$ the expectation
  value of the radial part of the matrix element.}\BibitemShut {Stop}%
\bibitem [{\citenamefont {Brookes}\ \emph {et~al.}(2018)\citenamefont
  {Brookes}, \citenamefont {Yakhou-Harris}, \citenamefont {Kummer},
  \citenamefont {Fondacaro}, \citenamefont {Cezar}, \citenamefont {Betto},
  \citenamefont {Velez-Fort}, \citenamefont {Amorese}, \citenamefont
  {Ghiringhelli}, \citenamefont {Braicovich}, \citenamefont {Barrett},
  \citenamefont {Berruyer}, \citenamefont {Cianciosi}, \citenamefont {Eybert},
  \citenamefont {Marion}, \citenamefont {van~der Linden},\ and\ \citenamefont
  {Zhang}}]{Brookes2018}%
  \BibitemOpen
  \bibfield  {author} {\bibinfo {author} {\bibfnamefont {N.B.}\ \bibnamefont
  {Brookes}}, \bibinfo {author} {\bibfnamefont {F.}~\bibnamefont
  {Yakhou-Harris}}, \bibinfo {author} {\bibfnamefont {K.}~\bibnamefont
  {Kummer}}, \bibinfo {author} {\bibfnamefont {A.}~\bibnamefont {Fondacaro}},
  \bibinfo {author} {\bibfnamefont {J.C.}\ \bibnamefont {Cezar}}, \bibinfo
  {author} {\bibfnamefont {D.}~\bibnamefont {Betto}}, \bibinfo {author}
  {\bibfnamefont {E.}~\bibnamefont {Velez-Fort}}, \bibinfo {author}
  {\bibfnamefont {A.}~\bibnamefont {Amorese}}, \bibinfo {author} {\bibfnamefont
  {G.}~\bibnamefont {Ghiringhelli}}, \bibinfo {author} {\bibfnamefont
  {L.}~\bibnamefont {Braicovich}}, \bibinfo {author} {\bibfnamefont
  {R.}~\bibnamefont {Barrett}}, \bibinfo {author} {\bibfnamefont
  {G.}~\bibnamefont {Berruyer}}, \bibinfo {author} {\bibfnamefont
  {F.}~\bibnamefont {Cianciosi}}, \bibinfo {author} {\bibfnamefont
  {L.}~\bibnamefont {Eybert}}, \bibinfo {author} {\bibfnamefont
  {P.}~\bibnamefont {Marion}}, \bibinfo {author} {\bibfnamefont
  {P.}~\bibnamefont {van~der Linden}}, \ and\ \bibinfo {author} {\bibfnamefont
  {L.}~\bibnamefont {Zhang}},\ }\bibfield  {title} {\enquote {\bibinfo {title}
  {The beamline {ID}32 at the {ESRF} for soft x-ray high energy resolution
  resonant inelastic x-ray scattering and polarisation dependent x-ray
  absorption spectroscopy},}\ }\href {\doibase
  https://doi.org/10.1016/j.nima.2018.07.001} {\bibfield  {journal} {\bibinfo
  {journal} {Nucl. Instrum. Methods Phys. Res. A}\ }\textbf {\bibinfo {volume}
  {903}},\ \bibinfo {pages} {175 -- 192} (\bibinfo {year} {2018})}\BibitemShut
  {NoStop}%
\bibitem [{\citenamefont {Haverkort}\ \emph {et~al.}(2012)\citenamefont
  {Haverkort}, \citenamefont {Zwierzycki},\ and\ \citenamefont
  {Andersen}}]{Haverkort2012}%
  \BibitemOpen
  \bibfield  {author} {\bibinfo {author} {\bibfnamefont {M.~W.}\ \bibnamefont
  {Haverkort}}, \bibinfo {author} {\bibfnamefont {M.}~\bibnamefont
  {Zwierzycki}}, \ and\ \bibinfo {author} {\bibfnamefont {O.~K.}\ \bibnamefont
  {Andersen}},\ }\bibfield  {title} {\enquote {\bibinfo {title} {Multiplet
  ligand-field theory using {W}annier orbitals},}\ }\href {\doibase
  10.1103/PhysRevB.85.165113} {\bibfield  {journal} {\bibinfo  {journal} {Phys.
  Rev. B}\ }\textbf {\bibinfo {volume} {85}},\ \bibinfo {pages} {165113}
  (\bibinfo {year} {2012})}\BibitemShut {NoStop}%
\bibitem [{\citenamefont {Haverkort}(2016)}]{Haverkort2016}%
  \BibitemOpen
  \bibfield  {author} {\bibinfo {author} {\bibfnamefont {M.~W.}\ \bibnamefont
  {Haverkort}},\ }\bibfield  {title} {\enquote {\bibinfo {title} {Quanty for
  core level spectroscopy - excitons, resonances and band excitations in time
  and frequency domain},}\ }\href
  {http://stacks.iop.org/1742-6596/712/i=1/a=012001} {\bibfield  {journal}
  {\bibinfo  {journal} {J. Phys.: Conf. Ser.}\ }\textbf {\bibinfo {volume}
  {712}},\ \bibinfo {pages} {012001} (\bibinfo {year} {2016})}\BibitemShut
  {NoStop}%
\bibitem [{\citenamefont {Cowan}(1981)}]{CowanBook}%
  \BibitemOpen
  \bibfield  {author} {\bibinfo {author} {\bibfnamefont {R.D.}\ \bibnamefont
  {Cowan}},\ }\href@noop {} {\emph {\bibinfo {title} {The theory of atomic
  structure and spectra.}}}\ (\bibinfo  {publisher} {University of California,
  Berkley},\ \bibinfo {year} {1981})\BibitemShut {NoStop}%
\bibitem [{\citenamefont {Sawatzky}\ and\ \citenamefont
  {Green}(2016)}]{Sawatzky}%
  \BibitemOpen
  \bibfield  {author} {\bibinfo {author} {\bibfnamefont {G.}~\bibnamefont
  {Sawatzky}}\ and\ \bibinfo {author} {\bibfnamefont {R.}~\bibnamefont
  {Green}},\ }\bibfield  {title} {\enquote {\bibinfo {title} {Quantum
  materials: Experiments and theory},}\ }\href
  {https://link.aps.org/doi/10.1103/PhysRevB.28.4315} {\bibfield  {journal}
  {\bibinfo  {journal} {edt. E. Pavarini, E. Koch, J. van den Brink and G.
  Sawatzky}\ }\textbf {\bibinfo {volume} {(Forschungszentrum Julich)}},\
  \bibinfo {pages} {1.1} (\bibinfo {year} {2016})}\BibitemShut {NoStop}%
\end{thebibliography}
\end{document}